\documentclass[11pt,letterpaper,nofootinbib]{revtex4-2}
\usepackage[top=100pt,bottom=0pt,letterpaper]{geometry}
\usepackage{adjustbox}
\usepackage{caption}
\usepackage[T1]{fontenc}
\usepackage{lmodern}
\usepackage[utf8]{inputenc}
\usepackage{microtype}
\usepackage{setspace}
\usepackage{float}
\restylefloat{table}
\usepackage{xcolor}
\usepackage{dcolumn}
\usepackage{lipsum}
\usepackage{graphicx,epsfig}
\usepackage{psfrag}
\usepackage{bm}
\usepackage{epstopdf}
\usepackage{latexsym}
\usepackage{multirow}
\usepackage{amsmath,amssymb}
\usepackage{enumerate}
\usepackage{hyperref}
\usepackage[FIGTOPCAP]{subfigure}
\def\be {\begin{equation}}
\def\ee {\end{equation}}
\def\ba {\begin{eqnarray}}
\def\ea {\end{eqnarray}}

\def\bc {\begin{center}}
\def\ec {\end{center}}
\newcommand{\bdm}{\begin{displaymath}}
\newcommand{\edm}{\end{displaymath}}

\def\a  {\alpha}
\def\b  {\beta}

\def\m  {\mu}

\def\th {\theta}

\def\ra {\rightarrow}

\def\la {\label}
\def\le {\left}
\def\ri {\right}

\def\f {\frac}
\def\sq {\sqrt}

\def\bi {\begin{itemize}}
\def\ei {\end{itemize}}

\def\> {\rangle}
\def\< {\langle}

\def\bc {\begin{center}}
\def\ec {\end{center}}

\begin{document}
\title{Generalised Ellis-Bronnikov wormholes embedded in warped braneworld background and energy conditions}

\author{Vivek Sharma} \email[email: ]{svivek829@gmail.com}
\author{Suman Ghosh} \email[email: ]{suman.ghosh@igntu.ac.in}
\affiliation{Department of Physics, Indira Gandhi National Tribal University, \\ Amarkantak, M.P. -484887,  India}

\begin{abstract}
 
Ellis-Bronnikov (EB) wormholes require violation of null energy conditions at the `throat'. This problem was cured by a simple modification of the `shape function', which introduces a new parameter $m\ge 2$ ($m=2$ corresponds to the EB model). This leads to a generalised (GEB) version. In this work, we consider a model where the GEB wormhole geometry is embedded in a five dimensional warped background. We studied the status of all the energy conditions in detail for both EB and GEB embedding. We present our results analytically (wherever possible) and graphically. Remarkably, the presence of decaying warp factor leads to satisfaction of weak energy conditions even for the EB geometry, while the status of all the other energy conditions are improved compared to the four dimensional scenario. Besides inventing a new way to avoid the presence of exotic matter, in order to form a wormhole passage, our work reveals yet another advantage of having a warped extra dimension.

\end{abstract}


\maketitle

\section{introduction}

Wormholes connect two distant spacetime points creating `short-cut's that allow `apparently faster than light' travel between those two points \cite{Visser-book,Hawking-book,Wald-book,Lobo-book,Witten:2019qhl}. 
Initially the idea was taken seriously by its inventors and proponents \cite{Flamm-1916,Einstein:1935tc,Fuller:1962zza}.
Soon it was realised that the Einstein-Rosen ``wormhole'' is not, contrary to expectations, a stable structure. The wormhole opens up and closes too quickly for even a photon to `travel' through.
Later work \cite{Morris:1988tu} suggested that exotic forms of energy threaded through a wormhole might keep it open  but it remains unclear whether such arrangements are physically feasible.
They have been analysed for many different reasons. However, the necessary presence of negative energy density makes them difficult to be created in macroscopic quantities \cite{Witten:2019qhl}. 
The original goal may have faded, but the Einstein-Rosen bridge still pops up occasionally as a handy solution to the pesky problem of intergalactic travel \cite{Morris:1988cz,Lobo:2017oab}.
Therefore, traversable wormholes are considered, by many, more of a science fiction rather than science. In classical general relativity, they are forbidden by the average null energy condition. 
Thus, within the framework of general relativity, wormholes are not real if energy conditions are to be absolutely respected \cite{Roman:2004xm, Hochberg:1998ha, Visser-book, Lobo-book}.
Note that, Quantum fluctuations permits the creation of microscopic wormholes \cite{Gao:2016bin,Maldacena:2018gjk,Fu:2018oaq,Fu:2019vco}. However, they are not adequate to form macroscopic wormholes. To be precise, they are allowed in the quantum theory, but, the time it takes to travel through the wormhole should be longer than the time it takes to travel between the two mouths on the outside. Therefore only microscopic wormholes were found using standard model matter.

However, it is also known for long that there are {\em classical} ways as well to circumvent the problem of exotic matter that violates the energy conditions \cite{Hochberg:1990is,Bhawal:1992sz,Agnese:1995kd,Samanta:2018hbw}. One such way is the framework of large class of the so-called modified theories of gravity. 
There are large number of such models exist in modified gravity that have {\em non-exotic} matter \cite{Lobo:2008zu,Kanti:2011jz,Kanti:2011yv,Zubair:2017oir,Shaikh:2016dpl,Ovgun:2018xys,Canate:2019spb}, though, in some cases, the convergence condition of null geodesics is violated. 
Models of dynamical wormholes \cite{Hochberg:1998ii,Roman:1992xj,Kar:1994tz,Kar:1995ss,Visser:2003yf} also provide ways of restricting the violation of energy conditions.
Other popular class modified gravity theories, where detailed analysis of wormhole geometries is done with viable matter source,  are the so-called $f(R)$ and higher order gravity theories \cite{Lobo:2009ip,Garcia:2010xb,MontelongoGarcia:2010xd,Sajadi:2012zz,Bertolami:2012fz,Harko:2013yb, Pavlovic:2014gba,Varieschi:2015wwa,Zangeneh:2015jda,Duplessis:2015xva,Mehdizadeh:2015jra,Samanta:2019tjb,Godani:2018blx,Rahaman:2014dpa,Rahaman:2016jds,Chakraborty:2021qwl,Mishra:2021ato}.
Recently, successful modelling and analysis of energy condition satisying wormholes are carried out within the framework of Born-Infeld gravity \cite{Maeda:2008nz,Shaikh:2015oha,Shaikh:2018yku} and torsional gravity \cite{Bohmer:2011si,Bronnikov:2015pha,Battista:2017hyj}.
Note that, all these different scenarios in general have different signatures in different physical phenomenon, particularly in gravitational lensing \cite{Cramer:1994qj,Dey:2008kn,Shaikh:2018oul}. 
With the advent of gravitational wave astronomy era \cite{Bishop:2021rye,Arimoto:2021cwc,Dent:2020nfa} and recent blackhole photography \cite{Akiyama:2019cqa}, search for wormhole signature do not seem unreal.
Recently, the possibility of existence of astrophysical wormholes in the dark matter galactic haloes is raised in \cite{Rahaman:2014pba,Rahaman:2013xoa}.
Wormholes have entered into the catalogue of the so-called blak hole mimickers and their unique signature can be imprinted on  during  merger phenomena \cite{Krishnendu:2017shb,Cardoso:2016oxy} or through nature of their quasi-normal modes \cite{Aneesh:2018hlp,Roy:2019yrr} etc. 
Such signature would also support the case for modified theories of gravity over general relativity.
The main purpose of this article is to investigate an yet unexplored modified gravity scenario namely a wormhole embedded in a warped five dimensional thick braneworld. Note that, earlier investigations are done on wormholes embedded in Kaluza-Klein, DGP and Randal-Sundrum thin braneworld scenario \cite{Lobo:2007qi,deLeon:2009pu,Wong:2011pt,Kar:2015lma,Banerjee:2019ssy,Wang:2017cnd} as such. Below we discuss motivation behind this analysis.

Though yet to be detected in experiments, extra dimensions are around in the literature for almost a century now \cite{Kaluza:1921tu,Klein:1926tv}. The reason behind the survival of this idea for so long is in the advantages one get in having them. For instance, while making models of unification (such as superstrings \cite{GSW}), or how the age-old hierarchy problem can be solved with extra dimensions \cite{Rubakov:1983bb,Gogberashvili:1998vx}. Recently, in the context of reinterpreting the standard model (and what may lye beyond) using octonions, extra dimensions appear {\em naturally} \cite{Furey:2015yxg}-\cite{Gillard:2019ygk}. Thus theories of extra dimensions have strong footing on basis of fundamental physical symmetries and are not mere useful extension of existing theories as such.
Perhaps the most popular among these higher dimensional models are the so-called warped braneworld models \cite{Gogberashvili:1998iu,Randall:1999ee,Randall:1999vf}. This model assumes a non-factorizable geometry-- a curved five dimensional spacetime where the geometry of the four-dimensional part depends on the extra dimension through a warping factor (a feature unique to this class of models).

Motivated by the appearance of extra dimensions in fundamental physics, we ask what new features extra dimensional models may induce on wormhole passages. Here we investigate a straightforward embedding of a four dimensional wormhole in a static five dimensional warped geometry. The family of wormholes we choose for embedding is based on \cite{Kar:1995jz} where the well-known Ellis-Bronnikov (EB) spacetime \cite{Ellis:1973yv,Bronnikov:1973fh} had been extended to provide a generalised family (GEB) of spacetimes that satisfies the null energy condition and further detailed studies is done in \cite{Roy:2019yrr}. For bulk geometry, we choose the so-called thick braneworld scenario \cite{Dzhunushaliev:2009va,Ghosh:2008vc} where the growing or decaying warping factor is a smooth function of the extra dimension and thus represent thick domain wall solutions. A thick brane scenario is preferred over the originally proposed infinitely thin Randall-Sundrum-branes as the former do not introduce Dirac delta functions in the field equations and they naturally appear if one takes into account quantum effects and minimum length scales.
Our intention here is to first figure out whether such models satisfy energy conditions or not, i.e., whether they admit matter sources that satisfy energy conditions. 
Recently it is reported that energy conditions are satisfied for wormholes embedded in Randall-Sundrum type thin brane models \cite{Sengupta:2021wvi}.
We on the other hand, want to see if the presence of a smooth warping factor can lead to a viable wormhole geometry with energy condition satisfying matter source. This is an yet unexplored feature of these class of models. 

Our program is as follows. In the next section, we briefly review the above mentioned generalised Ellis-Bronnikov wormhole geometry. Then we introduce our five dimensional model and the resulting field equations. In Section III,  we review (analytically wherever possible otherwise numerically/graphically) whether the energy conditions are satisfied (locally and/or globally) or not for the four dimensional model. Following this, we investigate how the status of the energy conditions are modified due to the presence of a warped extra dimension (with both decaying and growing warp factor). At the end we conclude with summarising the key results and future plans.



\section{Generalised Ellis--Bronnikov Wormhole and embedding in 5D warped Space-time}

Assuming phantom (negative kinetic energy) scalar field, Ellis and Bronnikov constructed static, spherically symmetric, geodesically complete, horizon-less Lorentzian wormhole-geometry connecting two asymptotically flat regions \cite{Ellis:1973yv,Bronnikov:1973fh}. The merric of the Ellis-Bronnikov wormhole is given by
\begin{equation}
ds^{2} = -dt^{2} + \f{dr^{2}}{1- \f{b_0^2}{r^2}} + r^{2}d\th^{2} + r^{2}\sin^{2}(\th)d\phi^{2} .\label{eq:E-B}
\end{equation}

Considering the freedom provided by the Morris-Thorne conditions \cite{Morris:1988cz} as necessary conditions to construct a Lorentzian wormhole, a generalised Ellis-Bronnikov (GEB) model is constructed by Kar et al. \cite{Kar:1995jz} as a two parameter ($m$ and $b_{0}$) family of Lorentzian wormholes, where $m$ is a free metric parameter and $b_{0}$ is the throat radius. For $m = 2$, one gets back the Ellis–Bronnikov spacetime.
The purpose behind the so-called GEB models was to study the their various classical properties such as the geodesics, the differences and similarities among such wormhole-geometries etc. The line element of GEB spacetime is given by
\begin{equation}
ds^{2} = -dt^{2} + dl^{2} + r^{2}(l)d\th^{2} + r^{2}(l)\sin^{2}(\th)d\phi^{2} ,\label{eq:EB-metric1}
\end{equation} 
\begin{equation}
\mbox{where} \hspace{2cm} r(l) = (b_{0}^{m} + l^{m})^{1/m} .\label{eq:r-l-relation}
 \end{equation}
The parameter $m$ takes only even values to make $r(l)$ smooth over the entire domain of the so-called `tortoise' or `proper radial distance' coordinate $l$~(where $-\infty$ $\leq$ $l$ $\leq$ $\infty$). Metric (\ref{eq:EB-metric1}), in terms of the usual radial coordinate $r$, can be written as
\begin{equation}
ds^{2} = -dt^{2} + \frac{dr^{2}}{(1 - \frac{b(r)}{r})} + r^{2}d\theta^{2} + r^{2}\sin^{2}\theta~ d\phi^{2} , \label{eq:EB-metric2}
\end{equation}
where $r$ and $l$ are related through the shape function $b(r)$ as,
\begin{equation}
dl^{2} = \frac{dr^{2}}{(1 - \frac{b(r)}{r})}, \hspace{1cm} b(r) = r - r^{(3 - 2m)} (r^{m} - b_{0}^{m})^{(2-\frac{2}{m})} . \label{eq:r&l-relation1}
\end{equation}

It is straightforward to derive the energy-momentum tensor that results in the geometry represented by metric (\ref{eq:EB-metric1}) using Einstein tensor and Einstein equations. In the frame basis (denoted by indices with hat), the diagonal components of the energy-momentum-tensor $T_{\bm\hat{\mu}\bm\hat{\nu}}~~ (\bm\hat{\mu}, \bm\hat{\nu} = \bm\hat{0}, \bm\hat{1}, \bm\hat{2}, \bm\hat{3})$ can be identified as $T_{\bm\hat{0}\bm\hat{0}} = \rho$, $T_{\bm\hat{1}\bm\hat{1}} = p_{1} = \tau$, $T_{\bm\hat{2}\bm\hat{2}} = p_{2}$ and $T_{\bm\hat{3}\bm\hat{3}} = p_{3}$,  where $\rho$ is the energy-density, $p_{1} = \tau$ is the radial tension, $p_{2}$ and $p_{3}$ are the principal pressure \cite{Visser-book} of the corresponding matter source. Due to spherical symmetry $p_{2}$ is equal to $p_{3}$. Thus the non-zero components of $T_{\bm\hat{\mu}\bm\hat{\nu}}$ for the GEB wormhole space-time (\ref{eq:EB-metric1}) are
\begin{equation}
T_{\bm\hat{0}\bm\hat{0}} = \rho(l) = - \frac{- 1 + r'^{2}(l) + 2 r(l) r''(l)}{r^{2}(l)}, \label{eq:EB-T_00}
\end{equation}
\begin{equation}
T_{\bm\hat{1}\bm\hat{1}} =p_{1}(l) = \tau(l) = \frac{- 1 + r'^{2}(l)}{r^{2}(l)} ,\label{eq:EB-T_11}
\end{equation}
\begin{equation}
T_{\bm\hat{2}\bm\hat{2}} = p_{2}(l) = \frac{r''(l)}{r(l)} ,\label{eq:EB-T_22}
\end{equation}
\begin{equation}
T_{\bm\hat{3}\bm\hat{3}} = p_{3}(l) = \frac{r''(l)}{r(l)} ,\label{eq:EB-T_33}
\end{equation}
where a prime denotes derivative with respect to $l$. In the following, we introduce a 5D warped spacetime where the 4D part is GEB wormhole. However the corresponding energy momentum tensor will be derived later.


A general warped line element in five dimensions is given as
\be
ds^2 = e^{2f(y)} g_{\a\b} dx^\a dx^\b + g_{44}dy^{2} \label{eq:RS}
\ee
where $g_{\a\b}$ can, in principle, be any metric and $g_{44}$ can be a function of 3-space, time, and the extra spatial dimension-$y$, not necessarily separable. The line element, representing an embedded wormhole, we choose to work with is as follows:
\begin{equation}
ds^2 = e^{2f(y)} \le(-dt^2 + dl^2 + r^{2}(l)d\theta^{2} + r^{2}(l)\sin^{2}\theta d\phi^{2} \ri) + dy^{2} . \label{eq:WGEB}
\end{equation}
Here the factor, $e^{2f(y)}$, is called a warp factor. The domain of $y$, which is the extra spatial dimension, can be $-\infty < y < \infty$. In the following, we shall set $f(y) = \pm\log[\cosh(y/y_0)]$, which correspond to the well-known thick brane models \cite{Dzhunushaliev:2009va,Ghosh:2008vc}. In such models, the brane is dynamically generated as a scalar field domain wall in the bulk. Note that the warp factor in such models is a smooth function of the extra dimension, unlike the Raldall-Sundrum models where $f(y) \sim |y|$ (i.e. a function with a derivative jump that implies presence of thin branes). Thick brane models do not posses the jumps and delta functions in the connection and curvature. They also appear naturally in multi-dimensional theories. In fact, Quantum fluctuations are expected to create an effective brane thickness. 
Note that the Ricci Scalar for metric (\ref{eq:WGEB}), is given by
\be
R_{5D} = 2e^{-2f}\le(\f{(2m-3)l^{2m-2}}{(b_0^m + l^m)^2} - \f{2(m-1)l^{m-2}}{b_0^m + l^m} + \f{1}{(b_0^m + l^m)^{2/m}}\ri) -4 (5f'^2 + 2 f'') .\la{eq:Ricci-5D}
\ee
Thus the curvature invariants of the warped model are essentially singularity free unlike some models of black holes in higher dimensions. However, if one considers thin brane models for embedding, then Dirac Delta functions would appear in the curvature to account for those infinitely thin branes.

\section{energy conditions}
The energy conditions (EC) are mathematical restriction on solutions of the Einstein equations, to rule out the {\em non-physical} solutions. A spacetime geometry may satisfy the one or many or all of the weak, strong, dominant and null energy conditions (WEC, SEC, DEC and NEC) to be physically viable solutions of Einstein equations. These energy conditions lead the following inequalities involving the energy and momentum densities--
\ba
 \rho &&\geq 0 , ~~~~ \rho + p_{i} \geq 0  ~~~~~~(\mbox{WEC}) , \la{eq:ec1}\\
 \rho + \sum_{i=1}^{3}p_{i} &&\geq 0 , ~~~~ \rho + p_{i} \geq 0 ~~~~~~(\mbox{SEC}) , \la{eq:ec2} \\
 \rho -   \mid p_{i} \mid &&\geq 0 ~~~~~~(\mbox{DEC}) , \la{eq:ec3} \\
\rho + p_{i} &&\geq 0 ~~~~~~(\mbox{NEC}) ~~~~~ \mbox{where $i = $ space index}.   \la{eq:ec4}
\ea 
Violation of these energy conditions would imply the existence of exotic matter (Matter with negative energy density). Note that, as the NEC is implied by the WEC, we would not discuss NEC in the present context. Primarily, Morris and Thorne found that for existence of stable {\em traversable} wormholes, violation of WEC is required atleast at the throat \cite{Visser-book}. Further studies revealed that presence of the so-called exotic matter is necessary for the stability of all classes of static wormholes. But, there are no observation in support for the presence of exotic matter, which raises doubt on the reality of wormholes.    \paragraph*{}

However, in various modified gravity models, as mentioned in the Introduction, modification of general relativity can serve the purpose of exotic matter. Thus providing stability to wormohole geometry even in presence of energy condition satisfying matter as such.
In the following we are going to explore whether the presence of warped extra dimension can lead to similar result.
To compare the two models given by Eq. (\ref{eq:EB-metric1}) and Eq. (\ref{eq:WGEB}), we  check analytically (wherever possible) and graphically how the energy conditions behaving throughout the spacetime. 

\subsection{Inequalities of WEC for GEB-Space-time}


WEC states that the energy density of any matter distribution should be non-negative for any time-like observer in space-time, which implies $\rho \geq 0$ and $\rho + p_{i} \geq 0$. 
Using Eqs (\ref{eq:EB-T_00}) to (\ref{eq:EB-T_33}), we define `inequality functions', $f^{(W)}_{\m}$ corresponding to WEC as
\begin{equation}
f^{(W)}_{0}(l)=\rho(l)=- \frac{- 1 + r'^{2}(l) + 2 r(l) r''(l)}{r^{2}(l)} ,\label{eq:EB-wec-f_0}
\end{equation} 
\begin{equation}
f_{1}^{(W)}(l) = \rho(l) + \tau(l) = - \frac{2r''(l)}{r(l)} ,\label{eq:EB-wec-f_1}
\end{equation}
\begin{equation}
f_{2}^{(W)}(l) = f_{3}^{(W)}(l) = \rho(l) + p_{2}(l) = - \frac{- 1 + r'(l)^{2} + r(l)r''(l)}{r(l)^{2}} . \label{eq:EB-wec-f_2}
\end{equation}
Here, prime denotes derivative with respect to the `Tortoise' coordinate $l$ and $f_{2}^{(W)} = f_{3}^{(W)} $ comes from the fact that $p_{2} = p_{3}$ (see Eqs. (\ref{eq:EB-T_22}) and (\ref{eq:EB-T_33})). According to the inequalities of WEC, the given functions ($f_{\mu}^{(W)}$, where $\mu = 0, 1, 2, 3$) should be greater than or equal to zero. Using the expression of $r(l)$ (given by Eq. (\ref{eq:r&l-relation1})) and setting $b_{0} = 1$, the inequality functions simplify as given in Table \ref{tab:tableW} for three cases with $m=2$, $m=4$ and $m=8$. 
\begin{center}
\begin{table}[h]
\begin{tabular}{|c|c|c|c|}
\hline

Function & $m = 2$ &$m = 4$ & $m = 8$ \\ \hline

$f_{0}^{(W)}$ &$-\frac{1}{(1+l^2)^{2}}$& $ \frac{1}{\sqrt{1+l^{4}}} - \frac{l^{2}(6+l^{4})}{(1 + l^{4})^{2}} $ & $ \frac{1}{(1 + l^{8})^{1/4}} - \frac{l^{6} (14 + l^{8})}{(1 + l^{8})^{2}}$ \\ \hline

$f_{1}^{(W)}$ &$-\frac{2}{(1+l^2)^{2}}$& $-\frac{6 l^{2}}{(1 +l^{4})^{2}} $ & $-\frac{14 l^{6}}{(1 + l^{8})^{2}} $ \\ \hline

$f_{2}^{(W)} = f_{3}^{(W)}$ & $0$ & $\frac{1}{\sq{1 + l^4}} - \frac{l^{2}(3 + l^{4})}{(1 + l^{4})^{2}}$ & $\frac{1}{(1 + l^{8})^{1/4}} - \frac{l^{6} (7 + l^{8})}{(1 + l^{8})^{2}} $ \\ \hline

\end{tabular}

\caption{Inequality functions, $f_{\mu}^{(W)}$ for $m=2$, $m=4$ and $m=8$.}

\label{tab:tableW}

\end{table}

\end{center}

\vspace{-1.3cm}
Thus, for $m=2$ the inequalities are maximally violated at and around the throat. However, in case of $m = 4$, we see that $f_{0}^{(W)}$ and $f_{2}^{(W)}$  are {\em positive} in the restricted domain given by  $-0.41 \leq l \leq 0.41$ and $-0.62 \leq l \leq 0.62$ respectively. The same is true for $m = 8$ in the domain $-0.64 \leq l \leq 0.64$ and $-0.73 \leq l \leq 0.73$. The second inequality of WEC ($f_{1}^{(W)} \geq 0$) is always violated (for all the cases), since, $f_{1}^{(W)}$ is negative for all values of $l$. Thus the WEC is partially satisfied so to speak.
We plot the inequality functions versus $l$ in Fig. \ref{fig:EB-WEC}. 
As $f_0^{(W)}$ is essentially equal to the matter energy density, the first plot in Fig. \ref{fig:EB-WEC} shows that the negative energy matter accumulates most near the throat for $m=2$ and moves away from the throat for $m>2$. Thus exotic matter gets localised inside a increasingly narrow region with increasing $m$. Thus giving a physical understanding of the parameter $m$. It can be further shown that for increasing $m$ the minima of the energy densities approaches $l = \pm b_0$. Thus characterising $\pm b_0$ as a length scale where there is uniformly distributed (in $l$) positive density matter bounded by two infinitely thin (in limit $m \ra \infty$) negative energy "walls", beyond which the of the energy density vanishes rapidly. 
\begin{figure}[H]
\centering
\includegraphics[scale=.4]{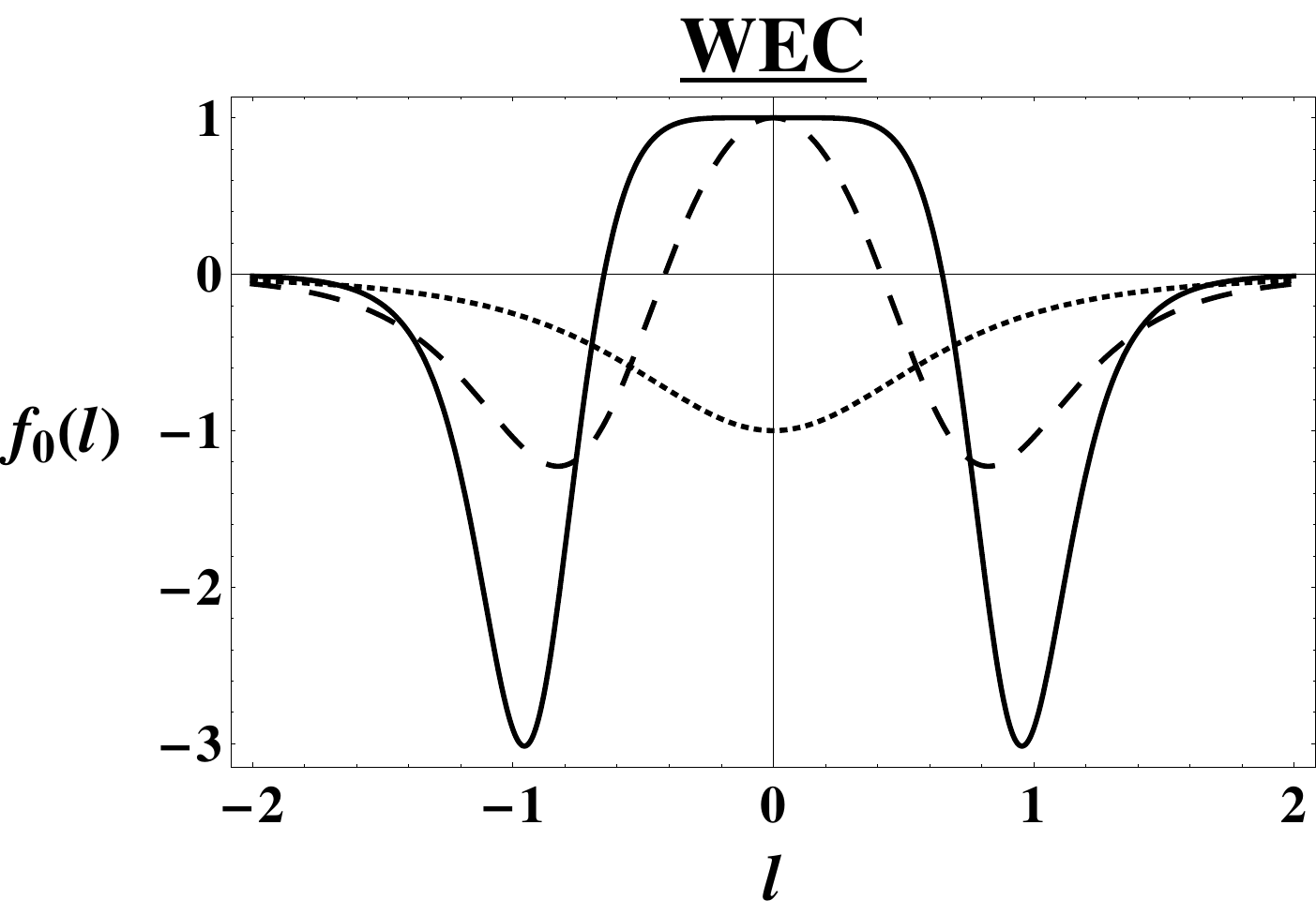}
\hspace{1cm} 
\includegraphics[scale=.4]{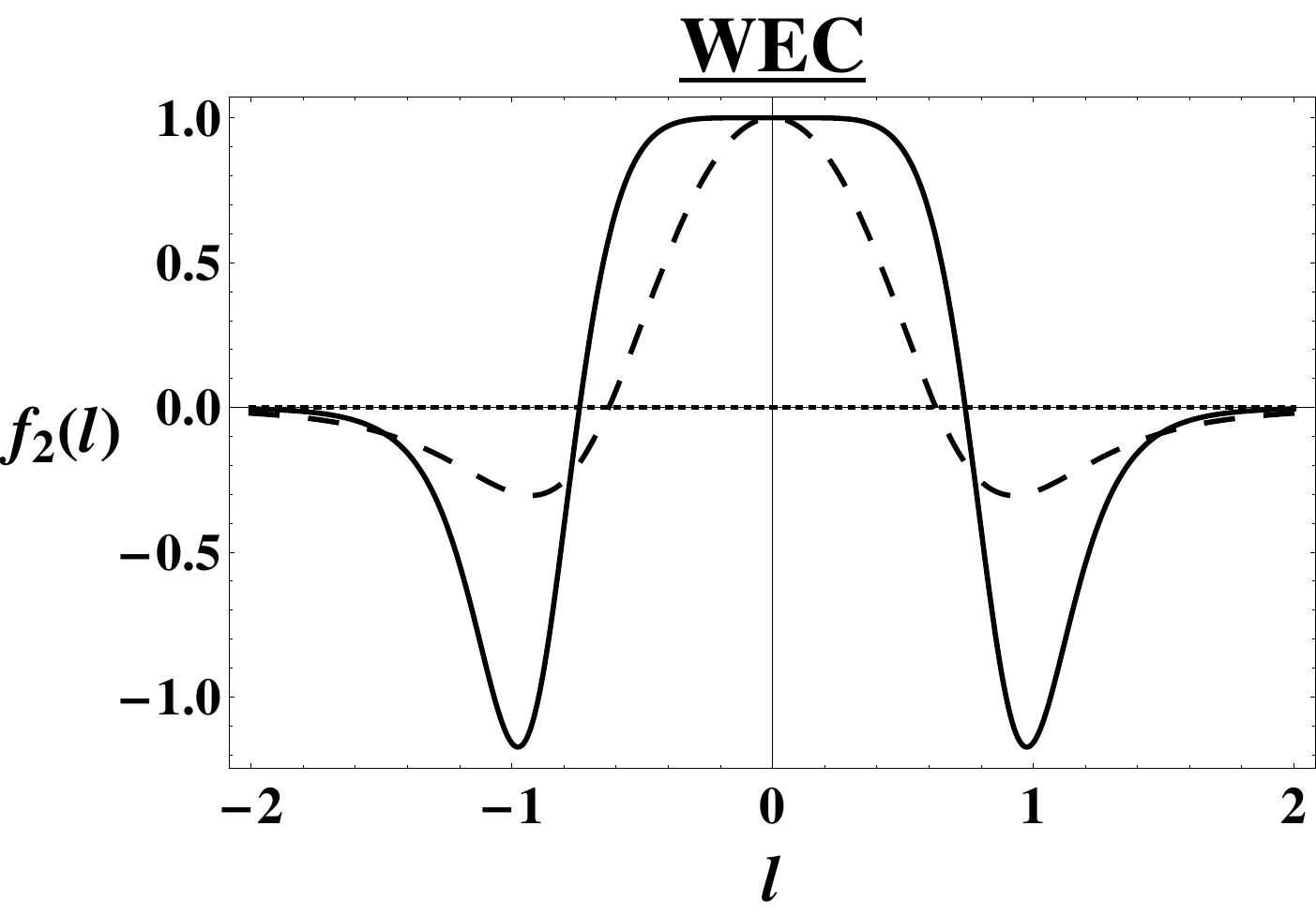}
\caption{Plot of inequality functions $f_{0}^{(W)}$ and $f_{2}^{(W)}$ (with $b_0 = 1$), where dotted curve represents case $m=2$, dashed curve represents case $m = 4$ and thick curve represents case $m = 8$.} 
\la{fig:EB-WEC}
\end{figure}




\subsection{Inequalities of SEC for GEB-Space-time}
SEC implies $\rho$ + $\displaystyle\sum_{j=1}^{3} p_j$ $\geq 0$, $\rho + p_{j} \geq 0$ ($j = 1, 2, 3$). Thus, from Eqs (\ref{eq:EB-T_00})-(\ref{eq:EB-T_33}) and (\ref{eq:EB-wec-f_1})-(\ref{eq:EB-wec-f_2}) we write the `inequality functions' for SEC, $f^{(S)}_{\m}$ as
\begin{equation}
f_{0}^{(S)}(l) = \rho(l) + \displaystyle\sum_{j=1}^{3} p_j(l) = 0 ,\label{eq:EB-sec_f0}
\end{equation} 
\begin{equation}
f_{1}^{(S)}(l) = f_{1}^{(W)}(l) ,\label{eq:EB-sec_f1}
\end{equation}
\begin{equation}
f_{2}^{(S)}(l) = f_{3}^{(S)}(l) = f_{2}^{(W)}(l) = f_{3}^{(W)}(l) .\label{eq:EB-sec_f2}
\end{equation}
Eq. (\ref{eq:EB-sec_f0}) implies that the first inequality function of SEC is always zero over the entire domain of coordinate $l$~($-\infty \leq l \leq \infty$). Eqs. (\ref{eq:EB-sec_f1}) and (\ref{eq:EB-sec_f2}), imply that the behaviour of these inequality functions are same as those analysed in the case of WEC.



\subsection{Inequalities of DEC for GEB-Space-time}

DEC essentially implies $\rho - \lvert p_{j} \rvert \geq 0$. Again, from Eqs (\ref{eq:EB-T_00}-\ref{eq:EB-T_33}) we write the corresponding inequality functions as
\begin{equation}
f_{1}^{(D)}(l) = \rho(l) - \lvert \tau(l)  \rvert = - \frac{- 1 + r'(l)^{2} + 2r(l)r''(l)}{r(l)^{2}} - \Big \lvert \frac{- 1 + r'(l)^{2}}{r(l)^{2}}  \Big \rvert ,\label{eq:EB-dec_f0}
\end{equation}
\begin{equation}
f_{2}^{(D)}(l) = f_{3}^{(D)}(l) = \rho(l) - \lvert p_{2}(l) \rvert = - \frac{- 1 + r'(l)^{2} + 2 r(l) r''(l)}{r(l)^{2}} - \Big \lvert \frac{r''(l)}{r(l)} \Big \rvert .\label{eq:EB-dec_f1}
\end{equation}
Again, by setting $b_{0} = 1$ and considering three cases with $m=2$, $m=4$ and $m=8$, the inequality functions $f_{\m}^{(D)}$ simplifies as given in Table \ref{tab:tableD}.
\begin{center}
\begin{adjustbox}{max width=\textwidth}
\begin{tabular}{|c|c|c|c|}
\hline
Function & $m = 2$ &$m = 4$ & $m = 8$ \\ \hline

$f_{1}^{(D)}(l)$ &$ - \frac{1}{(1+l^2)^{2}}  - \Big \lvert - \frac{1}{(1+l^2)^{2}}  \Big \rvert $ & $ \frac{1}{\sqrt{1 + l^{4}}} - \frac{l^{2} (6 + l^{4})}{(1 + l^{4})^{2}} - \Big \lvert \frac{l^{6}}{(1 + l^{4})^{2}} - \frac{1}{\sqrt{1 + l^{4}}} \Big \rvert  $ & $ \frac{1}{(1 + l^{8})^{1/4}} - \frac{l^{6} (14 + l^{8})}{(1 + l^{8})^{2}} - \Big \lvert \frac{l^{14}}{(1 + l^{8})^{2}} - \frac{1}{(1 + l^{8})^{1/4}} \Big \rvert $ \\ \hline

$f_{2}^{(D)}(l) = f_{3}^{(D)}(l)$ & $-\frac{1}{(1+l^2)^{2}} - \Big \lvert \frac{1}{(1+l^2)^{2}} \Big \rvert $ & $ \frac{1}{\sqrt{1 + l^{4}}} - \frac{l^{2} (6 + l^{4})}{(1 + l^{4})^{2}} - 3 \Big \lvert \frac{l^{2}}{(1 + l^{4})^{2}} \Big \rvert $ & $ \frac{1}{(1 + l^{8})^{1/4}} - \frac{l^{6} (14 + l^{8})}{(1 + l^{8})^{2}} - 7 \Big \lvert \frac{l^{6}}{(1 + l^{8})^{2}} \Big \rvert $ \\ \hline

\end{tabular}
\end{adjustbox}
\captionof{table}{Inequality functions, $f_{i}^{(D)}$  ($i=1, 2, 3$) for $m=2$, $m=4$ and $m=8$.}
\label{tab:tableD}
\end{center}
The two inequality functions $f_{1}^{(D)}$ and $f_{2}^{(D)}$ are plotted in Fig. \ref{fig:EB-DEC-f1} for three different values of $m$. In case of $m = 2$, DECs are violated everywhere in $l$. For $m = 4$, sum of the second and the third term in $f_{1}^{(D)}$ is greater than the first term for all values of $l$ (except for $l = 0$), therefore, $f_{1}^{(D)}$ will be always negative but vanishes at $l = 0$. Hence, first inequality of DEC is always violated (this is true for $m = 8$ as well). 
However, the second inequality of DEC is partially satisfied, since $f_{2}^{(D)}$ (or $f_{3}^{(D)}$) is positive for restricted domain of $l$ given by $\approx$ $-0.33 \leq l \leq 0.33$ (and  for $m = 8$, the domain is $-0.60 \leq l \leq 0.60$).

\begin{figure}[H]
\centering
\includegraphics[scale=.5]{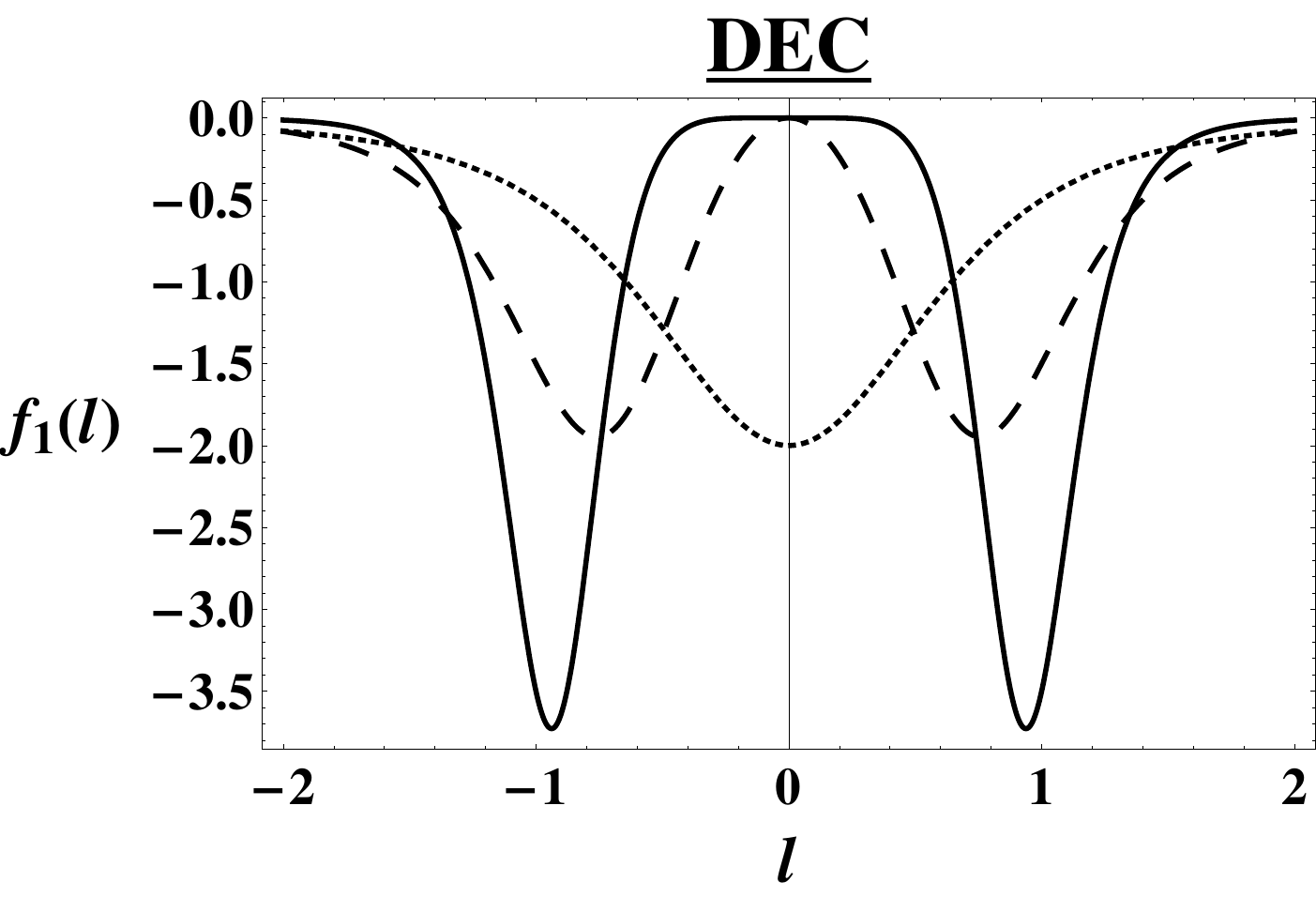}
\hspace{0.2cm}
\includegraphics[scale=.5]{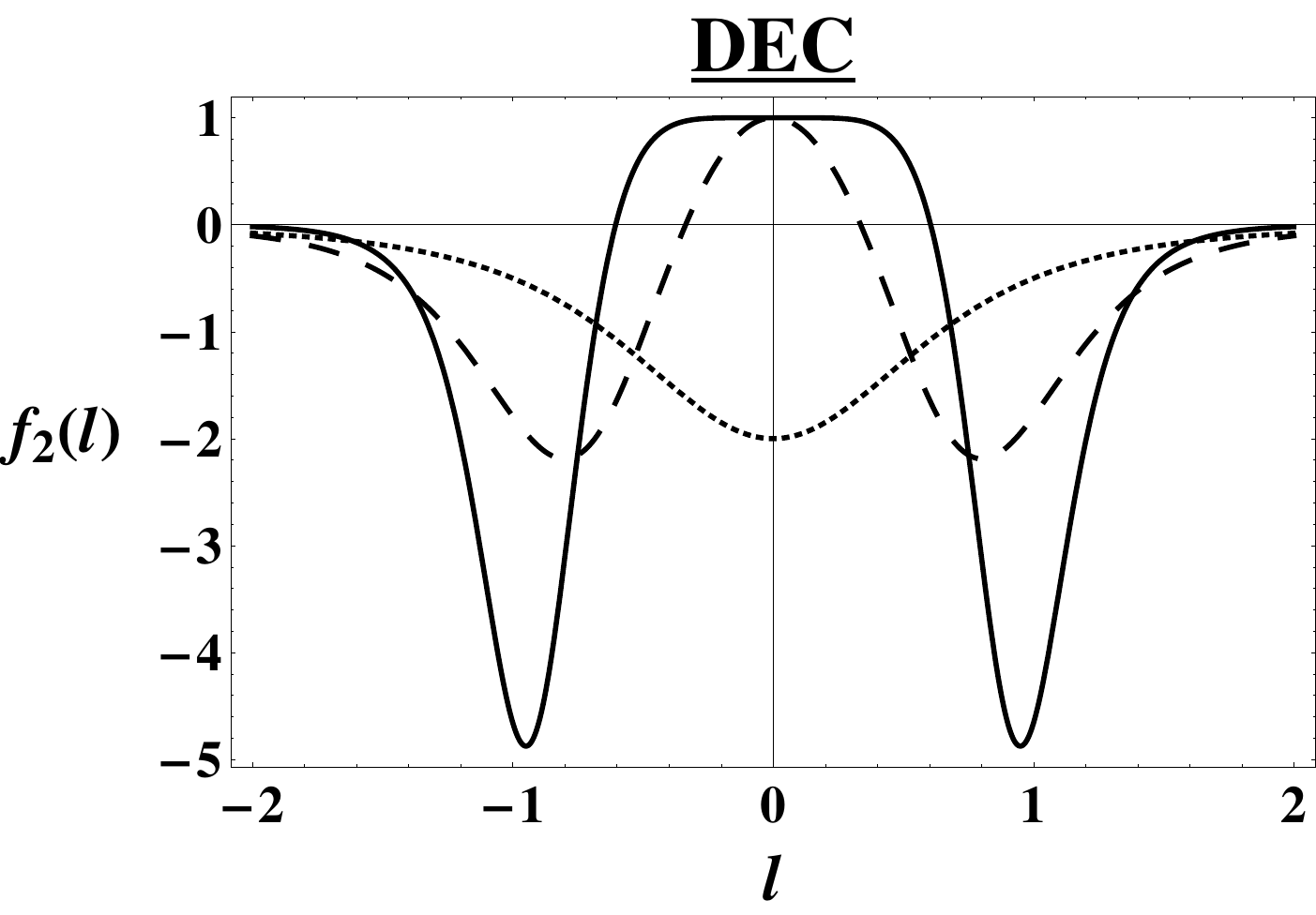} 
\caption{Plot of inequality functions of DEC (for $b_0 = 1$), where dotted curve represents case $m = 2$, dashed curve represents case $m = 4$ and thick curve represents the case $m = 8$  .} 
\la{fig:EB-DEC-f1}
\end{figure}

\begin{center}

\begin{adjustbox}{max width=\textwidth}
\begin{tabular}{|l|*{9}{c|}}
\hline
\multicolumn{1}{|l|}{GEB} & \multicolumn{1}{|l|}{} &   \multicolumn{3}{|c|}{$m=2$} & \multicolumn{1}{|l|}{}   & \multicolumn{3}{|c|}{$m>2$}\\ \hline

Ineq. Fns. & &   WEC   &   SEC  & DEC       & &  WEC   &   SEC   & DEC \\ \hline  \hline

$f_{0} \geq 0$ & & No & Yes  & --            & & Y   & Yes  & --  \\ \hline

$f_{1} \geq 0$ & & No & No & No             & & No  & No  & No  \\ \hline
 
$f_{2},f_{3} \geq 0$ & & Yes & Yes  & No    & & Y   & Y   & Y  \\ \hline
\end{tabular}
\end{adjustbox}
\captionof{table}{Status of energy conditions for GEB wormhole-geometry, where `Yes', `Y' and `No' stands for satisfied, partially satisfied and not satisfied respectively.}
\label{tab:EC-GEB}
\end{center}

Table \ref{tab:EC-GEB} summarises the status of GEB wormholes as per energy conditions are concerned. In case of $m=2$, the energy conditions are violated everywhere and maximally at the wormhole throat. However for $m \geq 2$ the energy conditions are partially satisfied near the throat. Thus GEB models provide a new way to avoid necessity of exotic matter. Let us briefly comment what this generalisation implies for $r(l)$. In case of $m=2$, $r(l)$ has a global minima at the throat ($l=0$) which corresponds to a global maxima in geodetic potential at the throat. Therefore the trapped trajectory denoted by $l=0$ is unstable. In case of $m>2$, on the other hand, $l=0$ is a saddle point for $r(l)$ (and also for the geodetic potential) and the non-vanishing and positively valued (and negative valued for geodetic potential) derivative of $r(l)$ appears at order $m$ \cite{Morris:1988cz}. Apparently, creating a saddle point in $r(l)$ is compensating for the negative energy density. 
After discussing energy conditions for GEB wormhole geometry in detail, in the following, we analyse the energy conditions for the GEB wormhole embedded in a warped background as given by Eq. (\ref{eq:WGEB}).

\section{Energy conditions for GEB spacetime in warped 5D background}

As mentioned earlier, we look to investigate whether the spacetime geometry given by Eq. (\ref{eq:WGEB}) may lead to satisfaction of energy conditions or not.
Eq. (\ref{eq:WGEB}) is a static and spherical symmetric geometry of spacetime in $(4+1)$ dimension, where $y$ is an extra spatial dimension. 
As usual we have, $r(l) = ({b_0}^{m} + l^{m})^{1/m}$. For the warp factor we choose $f(y) = \pm\log[\cosh(y/y_0)]$ that represents typical {\em thick} braneworld scenario with growing/decaying warp factor. Here, $y_0$ represents a characteristic length scale along the extra dimension.
For numerical computation we will set $b_0=1$ and $y_0=1$ in the following.
We get the energy-momentum tensor as the Einstein tensor for our metric (\ref{eq:WGEB}) as before. In the frame basis, the diagonal components of the energy-momentum-tensor are represented as $T_{\bm\hat{0}\bm\hat{0}} = \rho(l, y)$, $T_{\bm\hat{1}\bm\hat{1}} = p_{1}(l, y) = \tau(l, y)$, $T_{\bm\hat{2}\bm\hat{2}} = p_{2}(l, y)$, $T_{\bm\hat{3}\bm\hat{3}} = p_{3}(l, y)$, $T_{\bm\hat{4}\bm\hat{4}} = p_{4}(l, y)$ where $\rho$ is the energy-density, $p_{1} = \tau$ is the radial tension, $p_{2}$, $p_{3}$ and $p_{4}$ are the normal stresses in respective directions. Due to the spherical symmetry $p_{2}$ will be equal to $p_{3}$. Diagonal components of $T_{\bm\hat{\alpha}\bm\hat{\beta}}~~ (\bm\hat{\alpha}, \bm\hat{\beta} = 0, 1, 2, 3, 4)$ for our wormhole spacetime are given below.
\begin{equation}
T_{\bm\hat{0}\bm\hat{0}} = -3 \Big( 2 f'(y)^{2} + f''(y) \Big)-\frac{e^{-2f(y)} \Big( -1 + r'(l)^{2} + 2 r(l) r''(l) \Big)}{r(l)^{2}} ,\label{eq:t_00}
\end{equation}
\begin{equation}
T_{\bm\hat{1}\bm\hat{1}} = \frac{e^{-2f(y)} \Big( -1 + r'(l)^{2} \Big)}{r(l)^{2}} + 3 \Big( 2 f'(y)^{2} + f''(y)\Big) ,\label{eq:t_11}
\end{equation}
\begin{equation}
T_{\bm\hat{2}\bm\hat{2}} = 6 f'(y)^{2} + 3 f''(y) + \frac{e^{-2f(y)} r''(l)}{r(l)} ,\label{eq:t_22}
\end{equation}
\begin{equation}
T_{\bm\hat{3}\bm\hat{3}} = 6 f'(y)^{2} + 3 f''(y) + \frac{e^{-2f(y)} r''(l)}{r(l)} ,\label{eq:t_33}
\end{equation}
\begin{equation}
T_{\bm\hat{4}\bm\hat{4}} = 6 f'(y)^{2} + \frac{e^{-2f(y)} \Big( -1 + r'(l)^{2} + 2 r(l) r''(l)\Big)}{r(l)^{2}} ,\label{eq:t_44}
\end{equation}
Clearly, Eqs. (\ref{eq:t_00})-(\ref{eq:t_44}), suggests that the energy inequalities might behave differently due to the appearance of the new terms that depend on the derivatives of the decaying/growing warp factors. In the following, we analyse these inequality functions in detail.
 

\subsection{Inequalities of WEC}
The inequality functions $F_{A=01,2,3,4}$ for WEC corresponding to our 5D model (\ref{eq:WGEB}) are given below:
\begin{eqnarray}
F_{0}^{(W)}(l, y) &=& \rho(l, y)\nonumber \\
 &=& -3 \Big( 2 f'(y)^{2} + f''(y) \Big) - \frac{e^{-2f(y)} \Big( -1 + r'(l)^{2} + 2 r(l) r''(l) \Big)}{r(l)^{2}}  ,
\label{eq:WEC-F0}
\end{eqnarray}
\begin{equation}
F_{1}^{(W)}(l,y) = \rho(l, y) + \tau(l, y) = - \frac{2 e^{-2f(y)} r''(l)}{r(l)} ,
\label{eq:WEC-F1}
\end{equation}
\begin{eqnarray}
F_{2}^{(W)}(l,y) &=& F_{3}^{(W)}(l, y) = \rho(l, y) + p_{2}(l, y) \nonumber \\
&=& -\frac{e^{-2f(y)} \Big( -1 + r'(l)^{2} + r(l) r''(l) \Big)}{r(l)^{2}} ,\label{eq:WEC-F2}
\end{eqnarray}
\begin{equation}
F_{4}^{(W)}(y) = \rho(l, y) + p_{4}(l, y) = -3 f''(y) .
\label{eq:WEC-F4}
\end{equation}
It is easy to see that $F_1^{(W)}$ and $F_{2,3}^{(W)}$ only gets an overall positive multiplicative factor. So there status would not change compared to the 4D case. However, due to appearance of $y$-derivative terms, there is possibility that $F_0$ may satisfy the inequality (at least partially/locally) for decaying warp factor. It is also clear that, for growing warp factor there is no such hope. 
In the following, we analyse these inequalities as functions of $l$ at various locations along the extra dimension. In Table \ref{tab:WEC5}, for decaying warp factor, we present the analytical expressions of the inequality functions, at $y=0$ (typical location of the brane).
Note that, at $y=0$, $F_{1}^{(W)} = f_{1}^{(W)}$ and $F_{2}^{(W)} = F_{3}^{(W)} = f_{2}^{(W)} = f_{3}^{(W)}$, therefore behaviour of this inequalities everywhere in $l$ is similar as the four dimensional case. 
\begin{center}
\begin{table}[h]
\begin{tabular}{|c|c|c|c|}
\hline
Ineq. fn & $m=2$ & $m = 4$ & $m = 8$ \\ \hline
$F_{0}^{(W)}$ &$3-\frac{1}{(1+l^{2})^{2}}$ & $3 + \frac{1}{\sqrt{1+l^{4}}} - \frac{l^{2}(6+l^{4})}{(1 + l^{4})^{2}} $ & $3 + \frac{1}{(1 + l^{8})^{1/4}} - \frac{l^{6} (14 + l^{8})}{(1 + l^{8})^{2}}$ \\ \hline
$F_{1}^{(W)}$ &$-\frac{2}{(1+l^{2})^{2}}$ & $-\frac{6 l^{2}}{(1 +l^{4})^{2}} $ & $-\frac{14 l^{6}}{(1 + l^{8})^{2}} $ \\ \hline
$F_{2}^{(W)} = F_{3}^{(W)}$ & $0$ & $\frac{1}{(1 + l^{4})^{1/2}} - \frac{l^{2}(3 + l^{4})}{(1 + l^{4})^{2}}$ & $\frac{1}{(1 + l^{8})^{1/4}} - \frac{l^{6} (7 + l^{8})}{(1 + l^{8})^{2}} $ \\ \hline
$F_{4}^{(W)}$ & $3$ & $3$ & $3$ \\ \hline
\end{tabular}
\caption{Inequalities of WEC for decaying warp factor at $y = 0$ with $b_{0} = 1$, $m = 2, 4$ and $8$.} \label{tab:WEC5}
\end{table}
\end{center}
Remarkably, in contrast to the four dimensional case (see Table \ref{tab:tableW}), here we see that $F_0^{(W)} >0$ is satisfied for all values of $l$, even for $m=2$ (and for $m=4$ and partially for $m=8$) which is the most interesting result we report here. In fact, the whole energy density profile shifts towards the positive $F_0$ axis compared to the 4D case.
Fig. \ref{fig:5WEC-var} shows the variation of $F_{0}^{(W)}$ and $F_{2}^{(W)}$ Vs $l$ at $|y| = 0.4$ (variations are same for both signs of $y$). Again, for $m=2$ and $m=4$ energy density turns out to be positive everywhere in $l$. In fact, Eq. (\ref{eq:WEC-F2}) suggests, $F_2^{(W)}>0$ is satisfied everywhere on $y$ (for both decaying and growing warp factors) and the domain for $l$ is exactly same as that of 4D case.
We also do not need to plot $F_{1}^{(W)}$ (does not satisfy the inequality anywhere) and $F_{4}^{(W)}$ (does satisfy the inequality for decaying warp factor everywhere) vs $l$ as their behaviour is obvious from their analytic expressions.  

\begin{figure}[H]
\centering
\includegraphics[scale=.5]{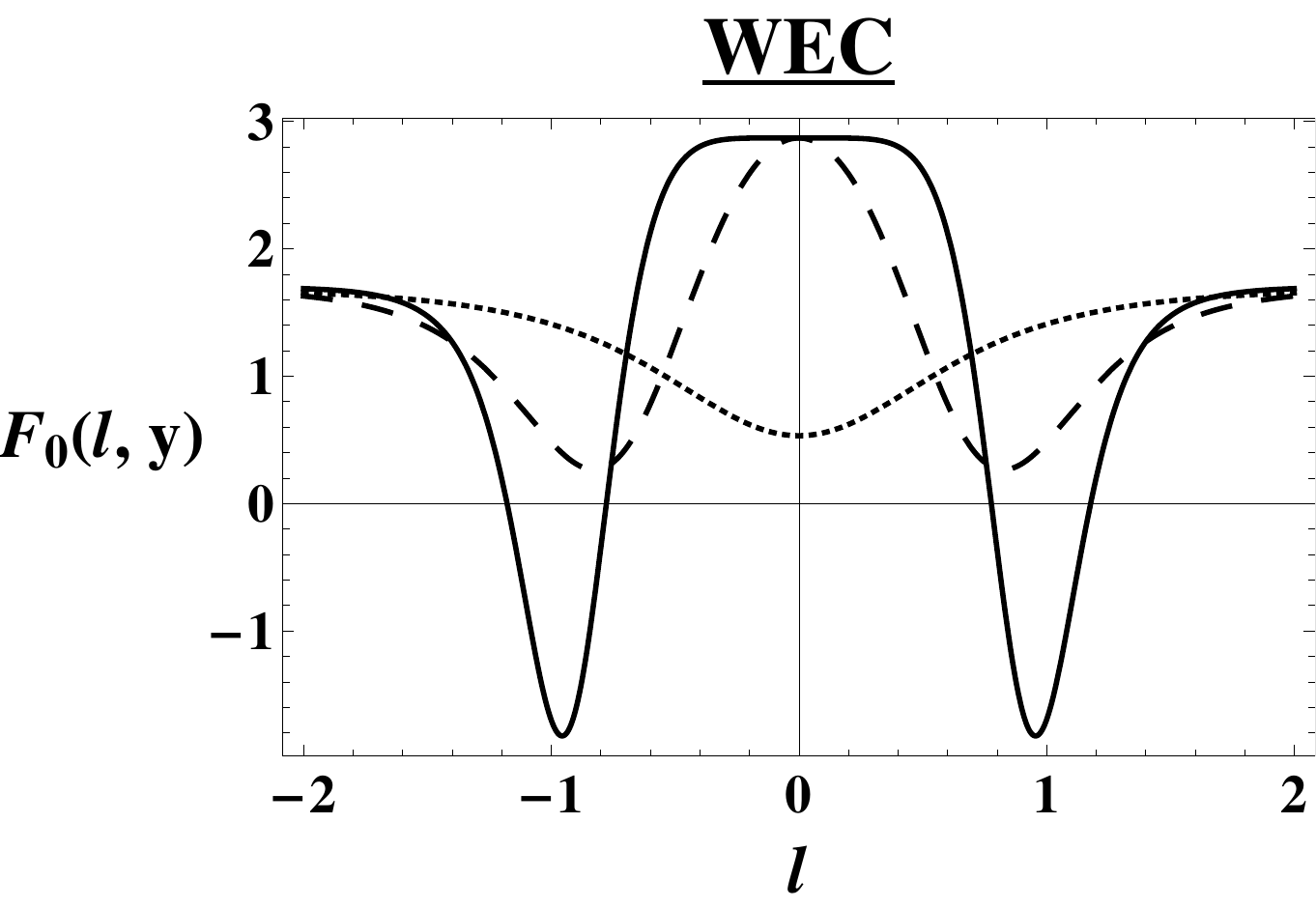}
\hspace{0.2cm}
 \includegraphics[scale=.5]{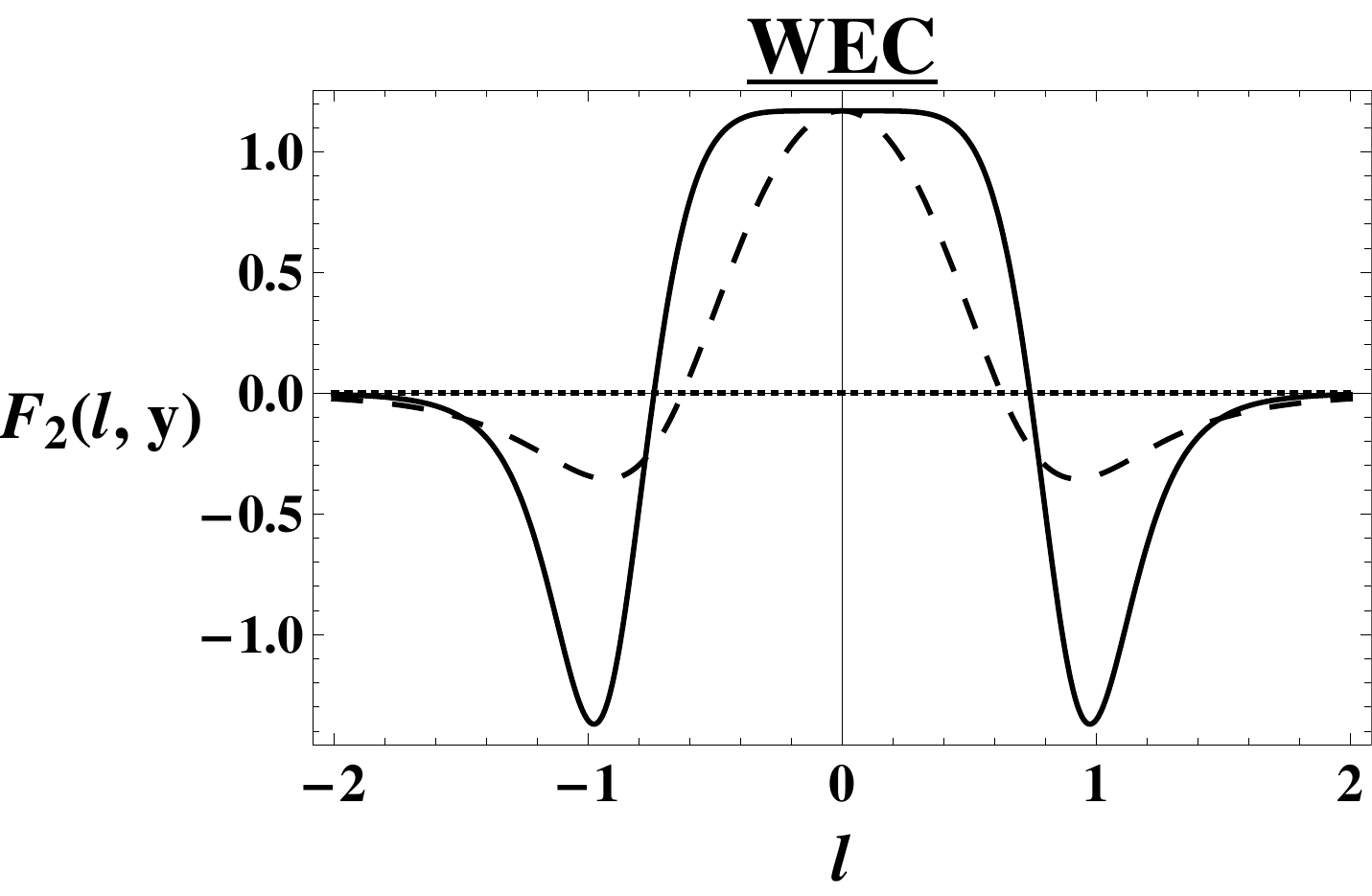}
\caption{Variation of inequality functions of WEC at $y = \pm 0.4$ (where $f(y) = -\log[\cosh(y)]$), with $b_0 = 1$ for cases-- $m = 2$ (dotted curve) $m = 4$ (dashed curve) and $m = 8$ (continuous curve).} \la{fig:5WEC-var}
\end{figure}

To get a complete view, we plot the energy density profile or $F_0^{(W)}$ in Fig. \ref{fig:5WEC-surfṭ} as a surface in the $l-y$ plane (for $m=2$ and $m=4$), which clearly shows that energy density is positive around $y=0$ and does become negative far away in the bulk. The physical reason behind satisfaction of WEC in presence of a decaying warp factor will be discussed at the end.

\begin{figure}[H]
\centering
\includegraphics[scale=.5]{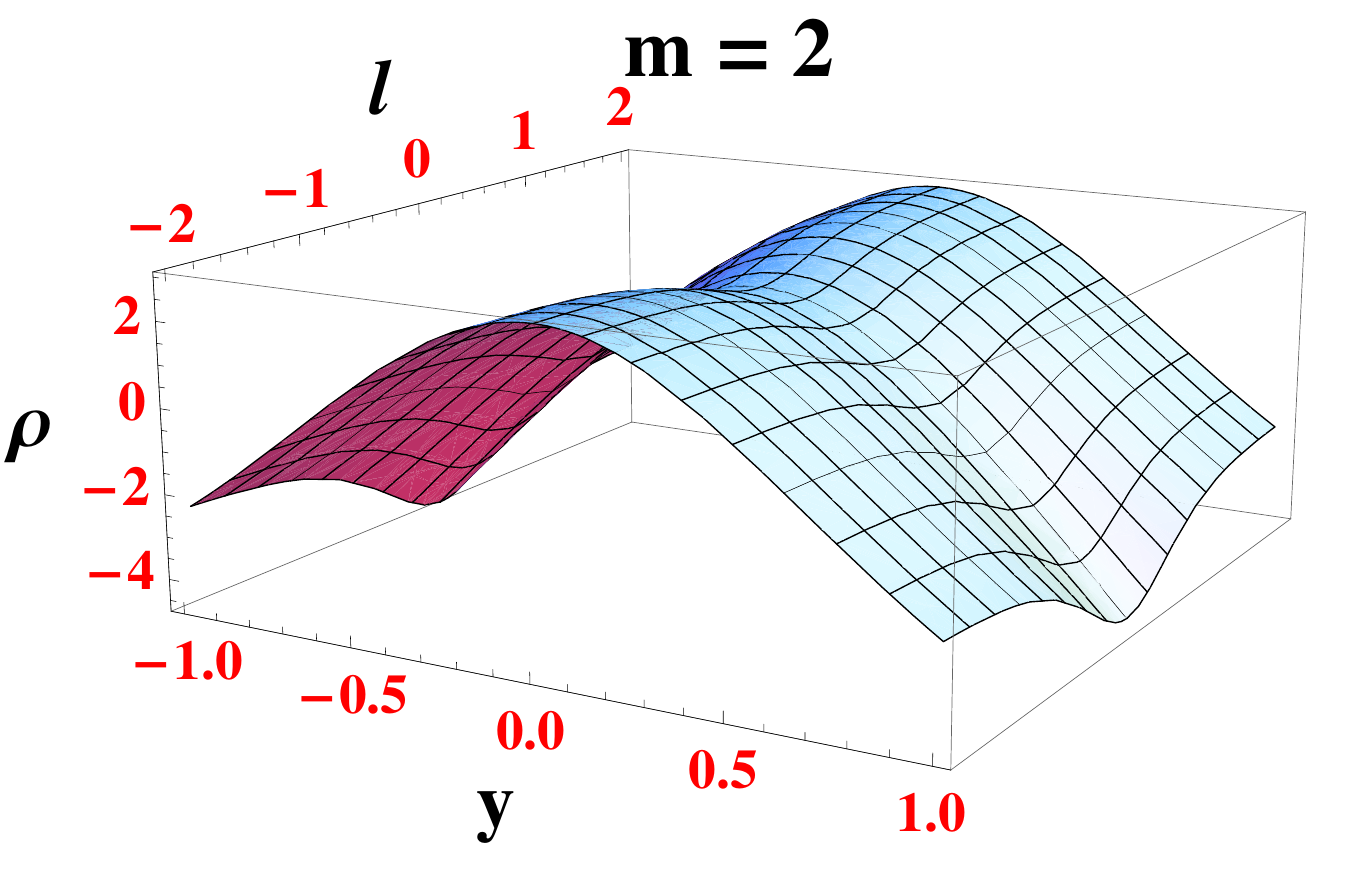}
\hspace{0.2cm}
 \includegraphics[scale=.51]{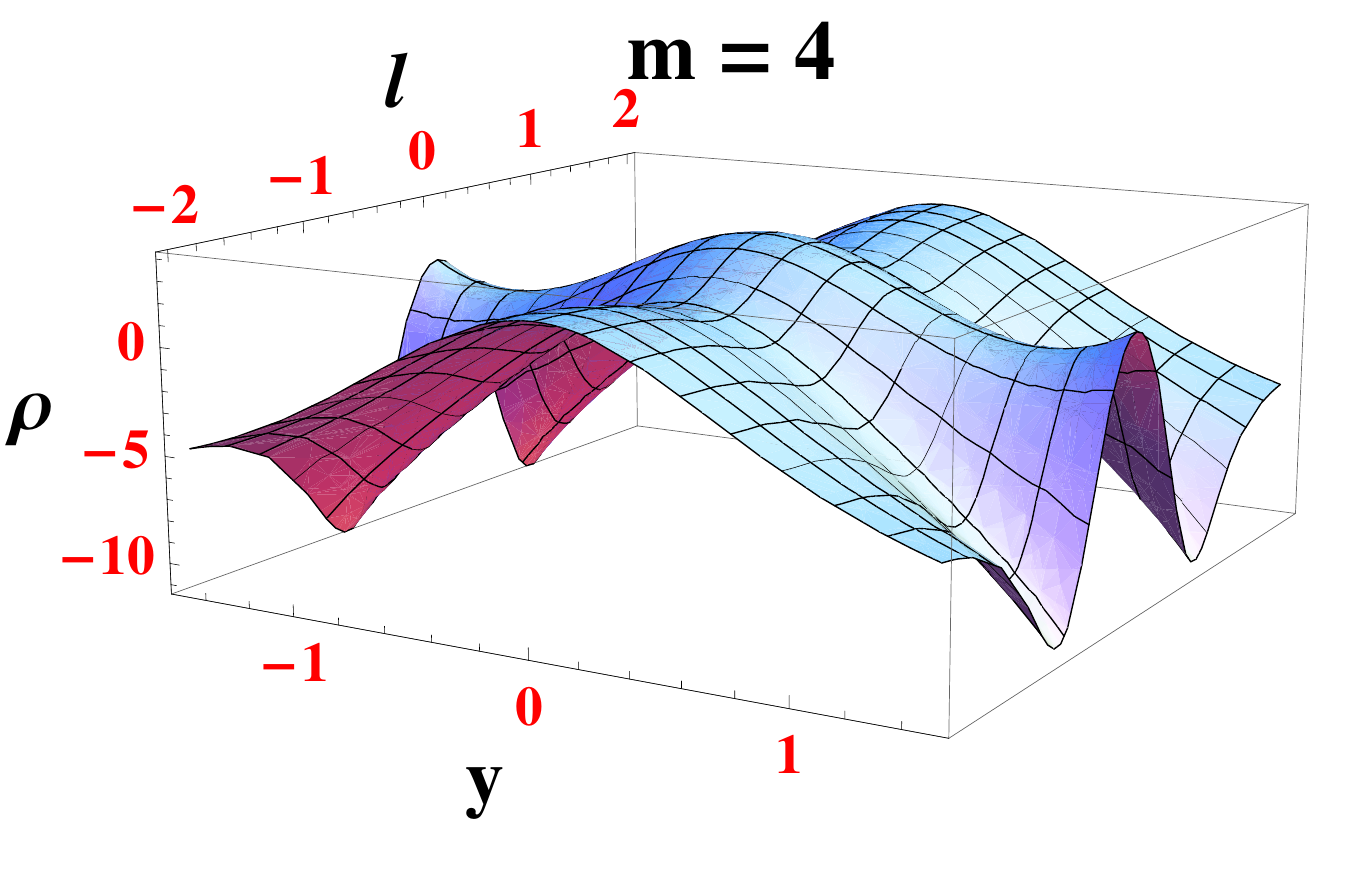} 
\caption{Variation of the energy density (where $f(y) = -\log[\cosh(y)]$), with $b_0 = 1$ for cases-- $m = 2$ and $m=4$.} \la{fig:5WEC-surfṭ}
\end{figure}


This prompts us to find the parameter space or the domain of the $l-y$ plane where  $F_0^{(W)} >0$ is satisfied in case of decaying warp factor. This is depicted in Fig \ref{fig:5WEC-f0}. The shaded region is the parameter space where the inequality is satisfied. It is noticeable from Fig \ref{fig:5WEC-f0} that for increasing $m$, the domain of $y$ is converging on $y=0$ given that the energy density is positive everywhere on $l$. This suggests then that, in the $m \ra \infty$ limit, thickness of a {\em physical} (made of positive energy density matter) thick brane vanishes.

\begin{figure}[H]
\centering
\includegraphics[scale=.27]{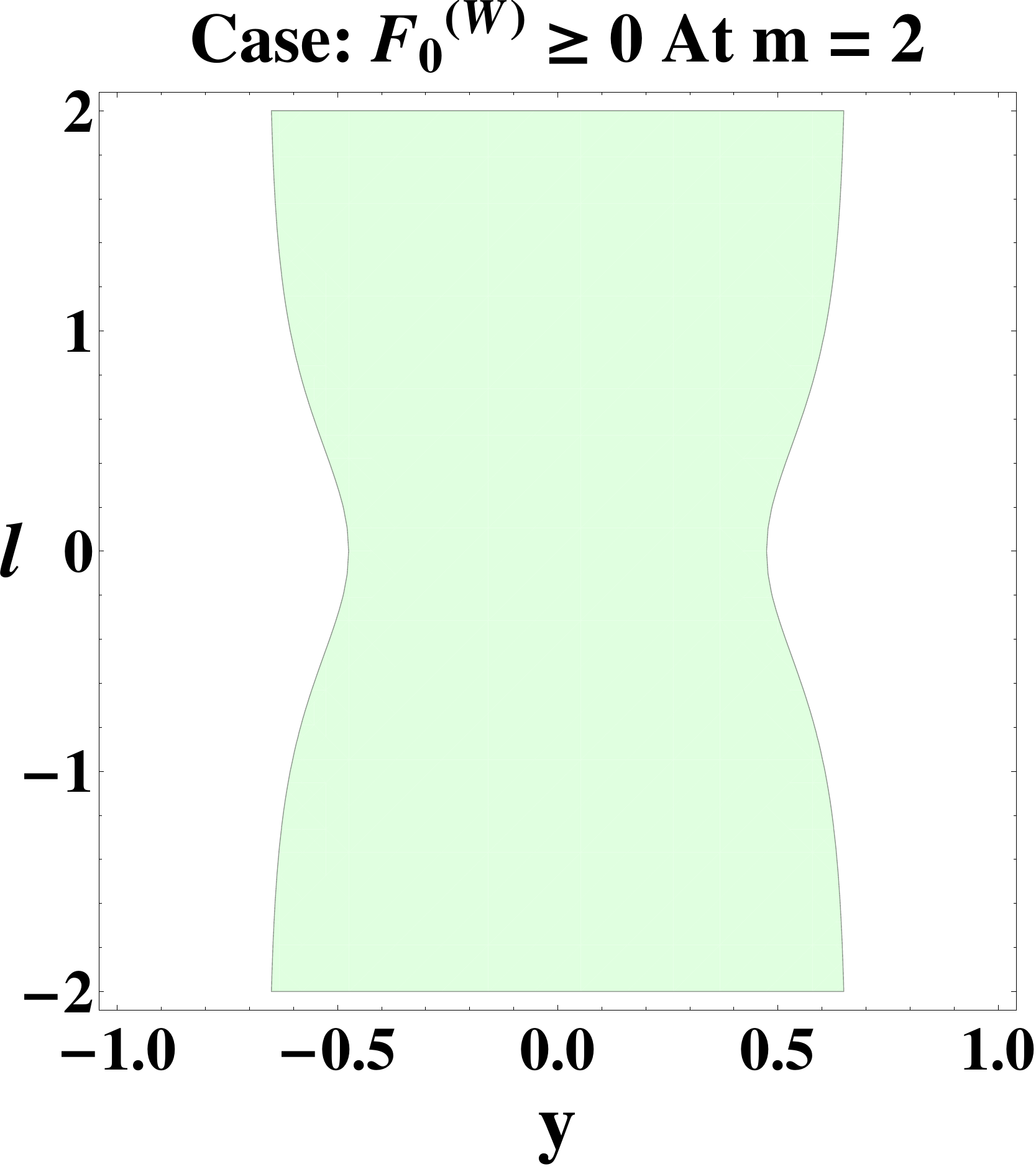}
\hspace{0.2cm}
 \includegraphics[scale=.265]{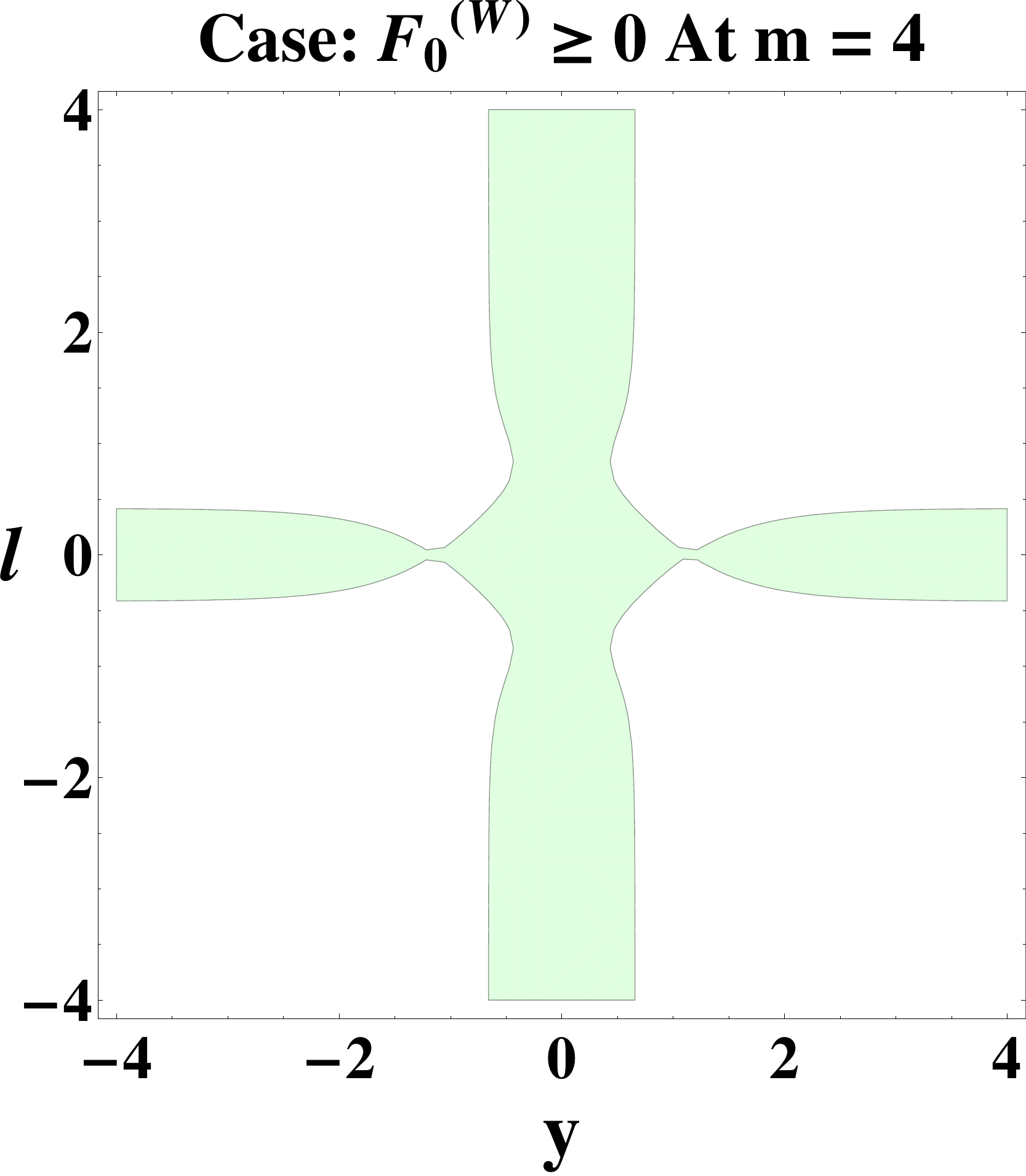}
 \hspace{0.2cm}
 \includegraphics[scale=.265]{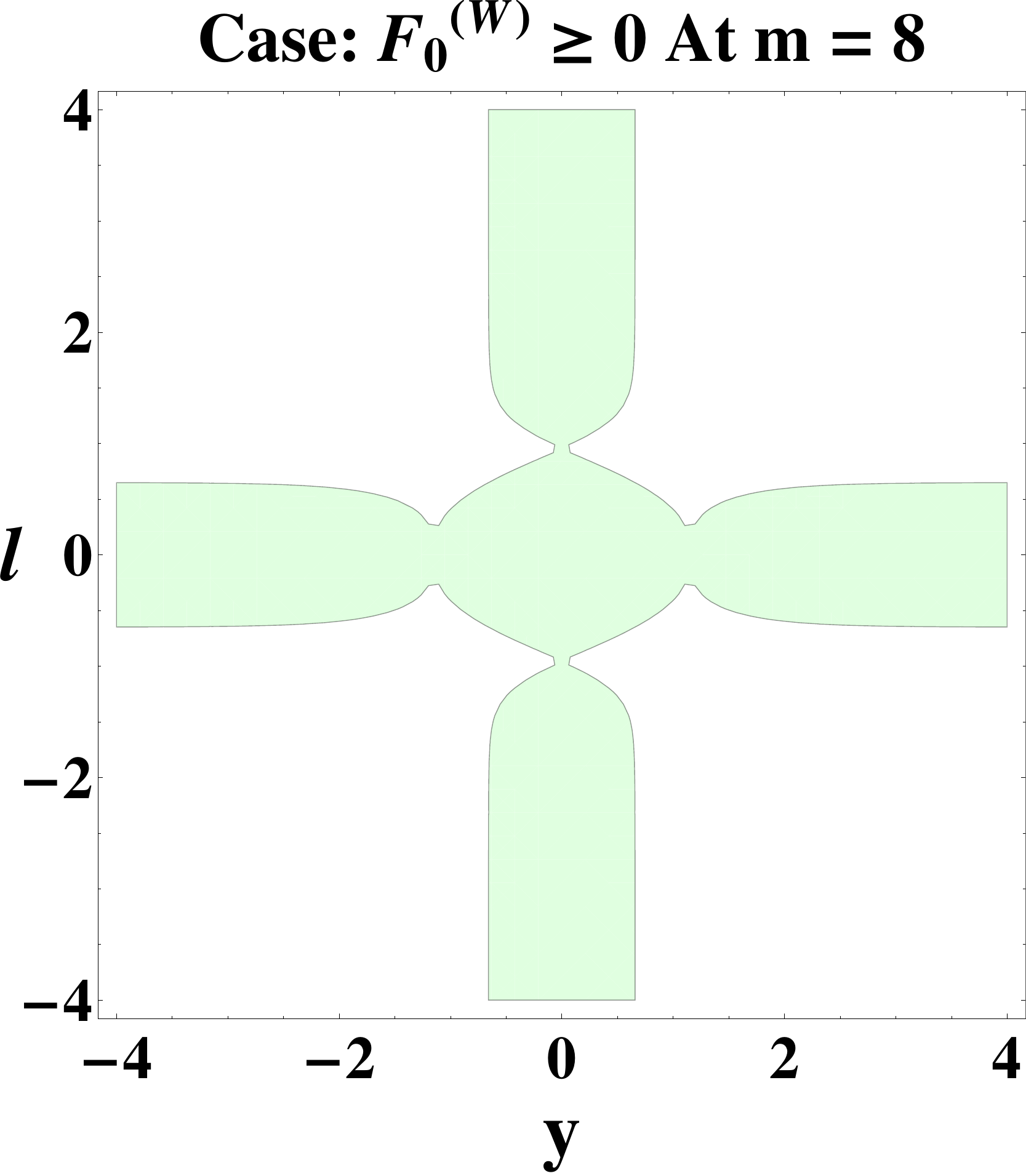}
 \caption{Parameter space plot $F_0^{(W)}>0$ (shaded region), with $f(y) = -\log[\cosh(y)]$ for $m = 2$ $m = 4$ and $m = 8$  with $b_0 = 1$.} \la{fig:5WEC-f0}
\end{figure}

\subsection{Inequalities of SEC}

As usual, we can write the first inequality, $\rho(l, y)+ \displaystyle\sum_{i=1}^{4} p_i(l, y) \geq 0$, of SEC  using Eqs (\ref{eq:t_00}) - (\ref{eq:t_44}). All remaining inequalities of SEC, $\rho(l, y) + p_{i}(l, y) \geq 0$, are already being implied in the WEC section above. The inequality function $F_{0}^{(S)}$ for SEC is,
\begin{equation}
F_{0}^{(S)}(l, y) = 6 \Big( 3f'(y)^{2} + f''(y)\Big) + \frac{e^{-2f(y)} \Big( -1 + r'(l)^{2} + 2 r(l)r''(l)\Big)}{r(l)^{2}} .\label{eq:SEC-F0}
\end{equation}
In Table \ref{tab:SEC5}, we write the analytic expression of $F_{0}^{(S)}$, for both decaying and growing warp factor, at $y=0$, as a function of $l$ (with $b_0=1$, $y_0=1$) and for values of $m = 2, 4, 8$. 
\begin{center}
\begin{adjustbox}{max width=\textwidth}
\begin{tabular}{|c|c|c|c|}
\hline

$F_{0}^{(S)}$ & $m = 2$ & $m = 4$ & $m = 8$ \\ \hline

Decaying & $-6+\frac{1}{(1+l^{2})^{2}}$ & $- 6 - \frac{1}{\sqrt{1 + l^{4}}} + \frac{l^{2} (6 + l^{4})}{(1 + l^{4})^{2}}$ & $- 6 - \frac{1}{(1 + l^{8})^{1/4}} + \frac{l^{6} (14 + l^{8})}{(1 + l^{8})^{2}}$ \\ \hline

Growing & $6+\frac{1}{(1+l^{2})^{2}}$ & $ 6 - \frac{1}{\sqrt{1 + l^{4}}} + \frac{l^{2} (6 + l^{4})}{(1 + l^{4})^{2}}$ & $6 - \frac{1}{(1 + l^{8})^{1/4}} + \frac{l^{6} (14 + l^{8})}{(1 + l^{8})^{2}}$ \\ \hline




\end{tabular}
\end{adjustbox}
\captionof{table}{$F_{0}^{(S)}$ for $f(y) = -\log[\cosh(y/y_0)]$ (first row) and $f(y) = +\log[\cosh(y/y_0)]$ (second row) at $y = 0$ with $b_{0} = 1$, $y_0=1$ and  $m =2, 4, 8$.}
\label{tab:SEC5}
\end{center}
Table \ref{tab:SEC5} implies that, for the case of decaying warp factor, at or near $y=0$, the inequality is not satisfied anywhere in $l$ for all $m$. On the other hand, this inequality is always satisfied everywhere in $l$ in the growing warp factor scenario.  
Let us to look at the parameter space again in the case of decaying warp factor which is presented in Fig. \ref{fig:5SEC-f0}. This shows that for $m=2$ the inequality is satisfied almost everywhere in $l-y$ plane except the region which essentially represents the location of the thick brane. However, for of $m\ra \infty$, the inequality is satisfied at or near of $l \ra \pm 1$ (these are the regions about to pinch for large $m$ in Fig. \ref{fig:5SEC-f0}). On the other hand, in presence of the growing warp factor, the inequality is satisfied everywhere in the $l-y$ plane as is obvious from Eq. (\ref{eq:SEC-F0}).

\begin{figure}[H]
\centering
\includegraphics[scale=.26]{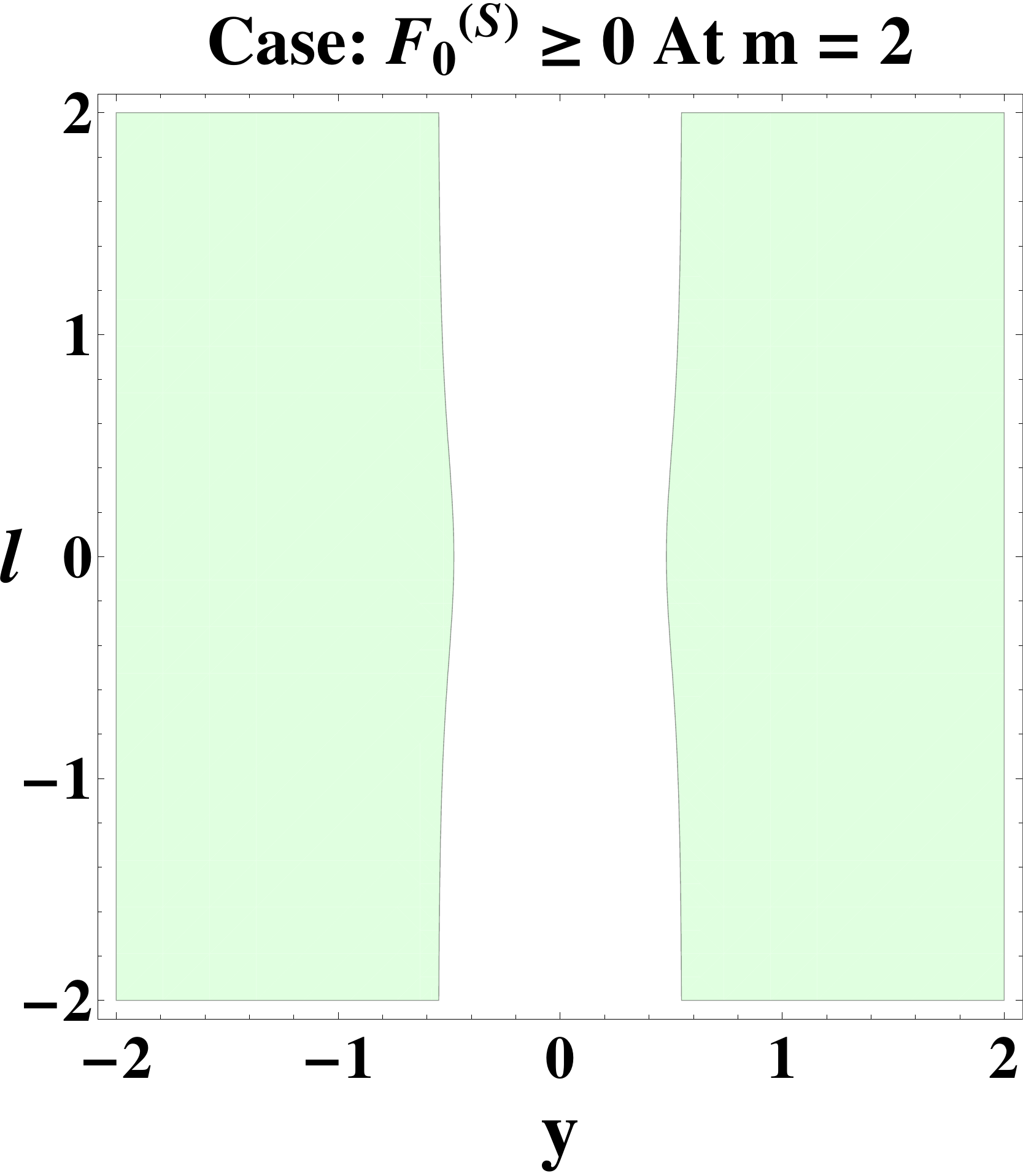}
\hspace{0.2cm}
\includegraphics[scale=.26]{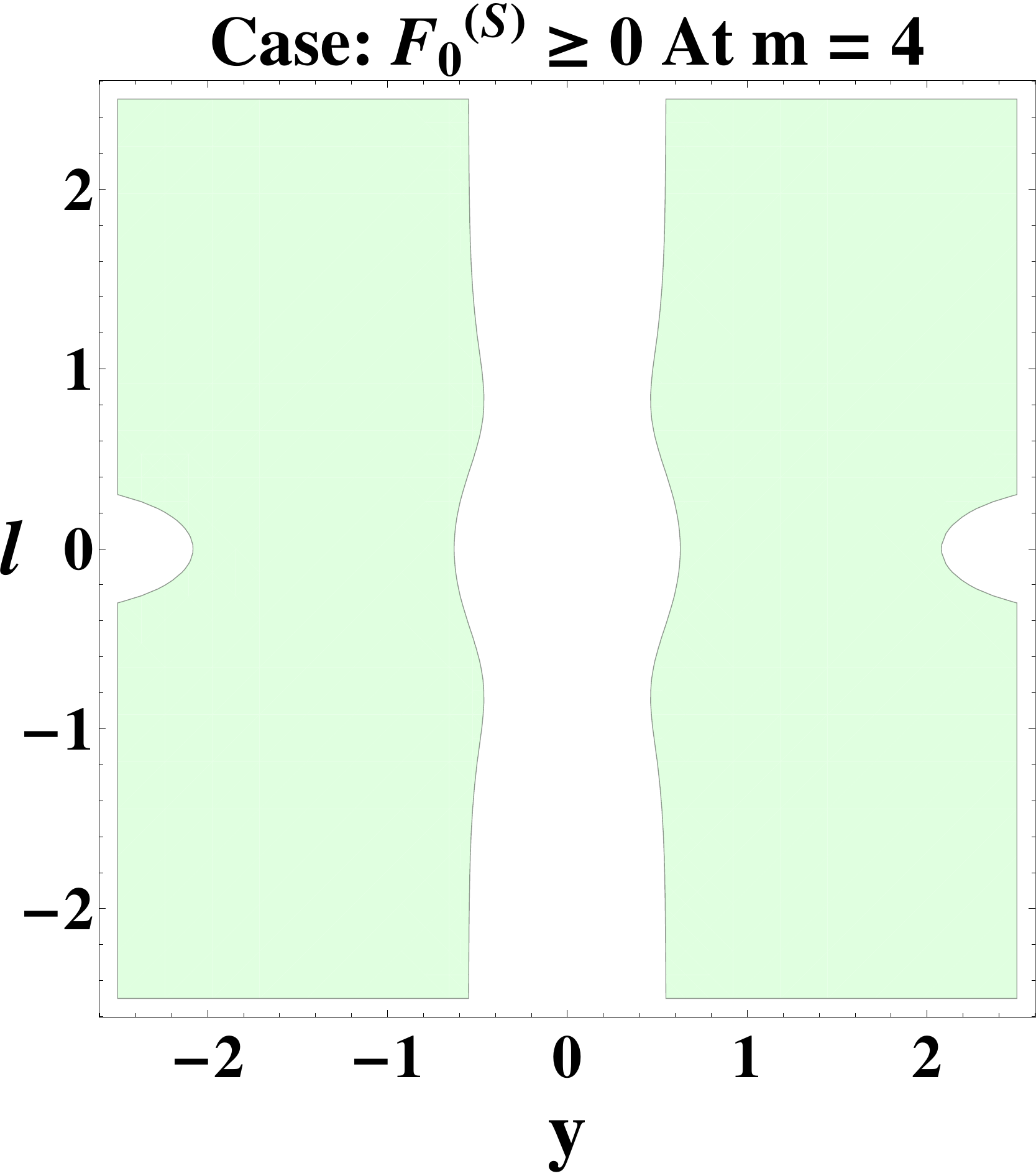}
 \hspace{0.2cm}
 \includegraphics[scale=.26]{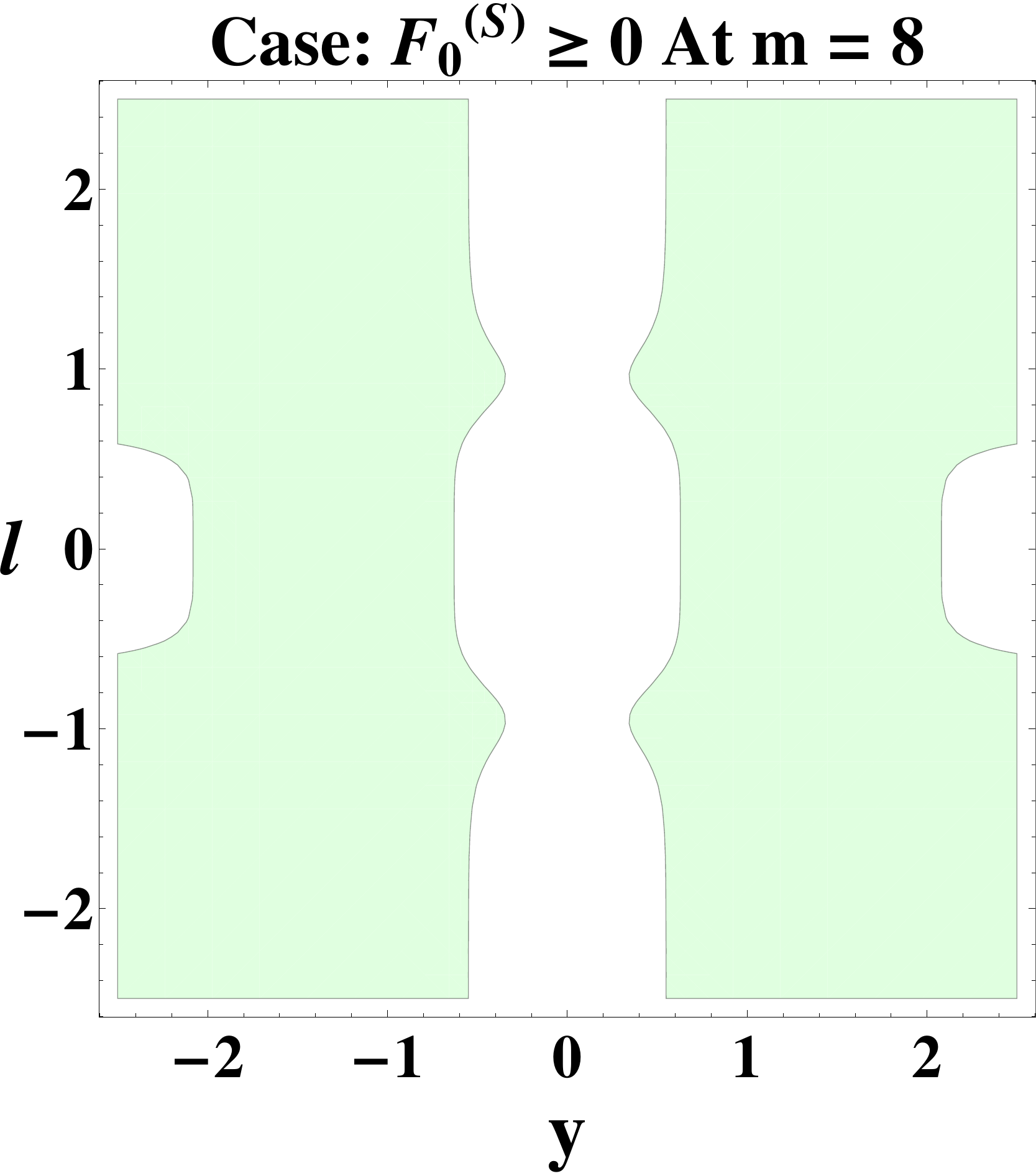} 
\caption{Parameter space plots for $F_{0}^{(S)}>0$ (where $f(y) = - \log[\cosh(y)]$) for $m = 2,4,8$.} \la{fig:5SEC-f0}
\end{figure}
Taking suggestion from the parameter space plots, in Fig. \ref{fig:SEC-f0}, we present the variation of $F_{0}(l, y)$ Vs $l$ at $|y| = 1$ (for $m=2,4,8$), where the inequality is satisfied in presence of both decaying and growing warp factor. The distinctive feature of $m=2$ case is again visible.

\begin{figure}[H]
\centering
\includegraphics[scale=.45]{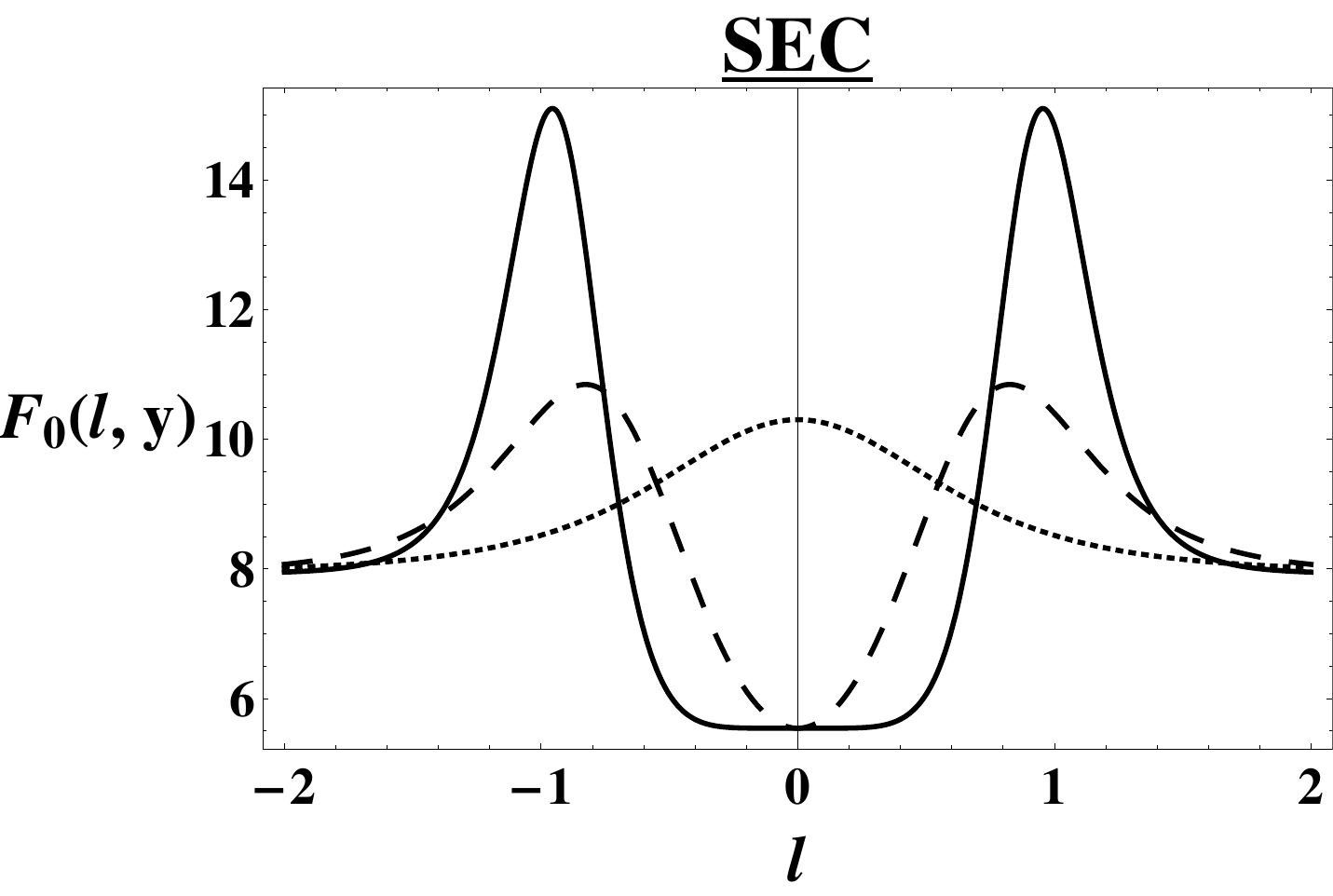}
\hspace{0.3cm}
\includegraphics[scale=.45]{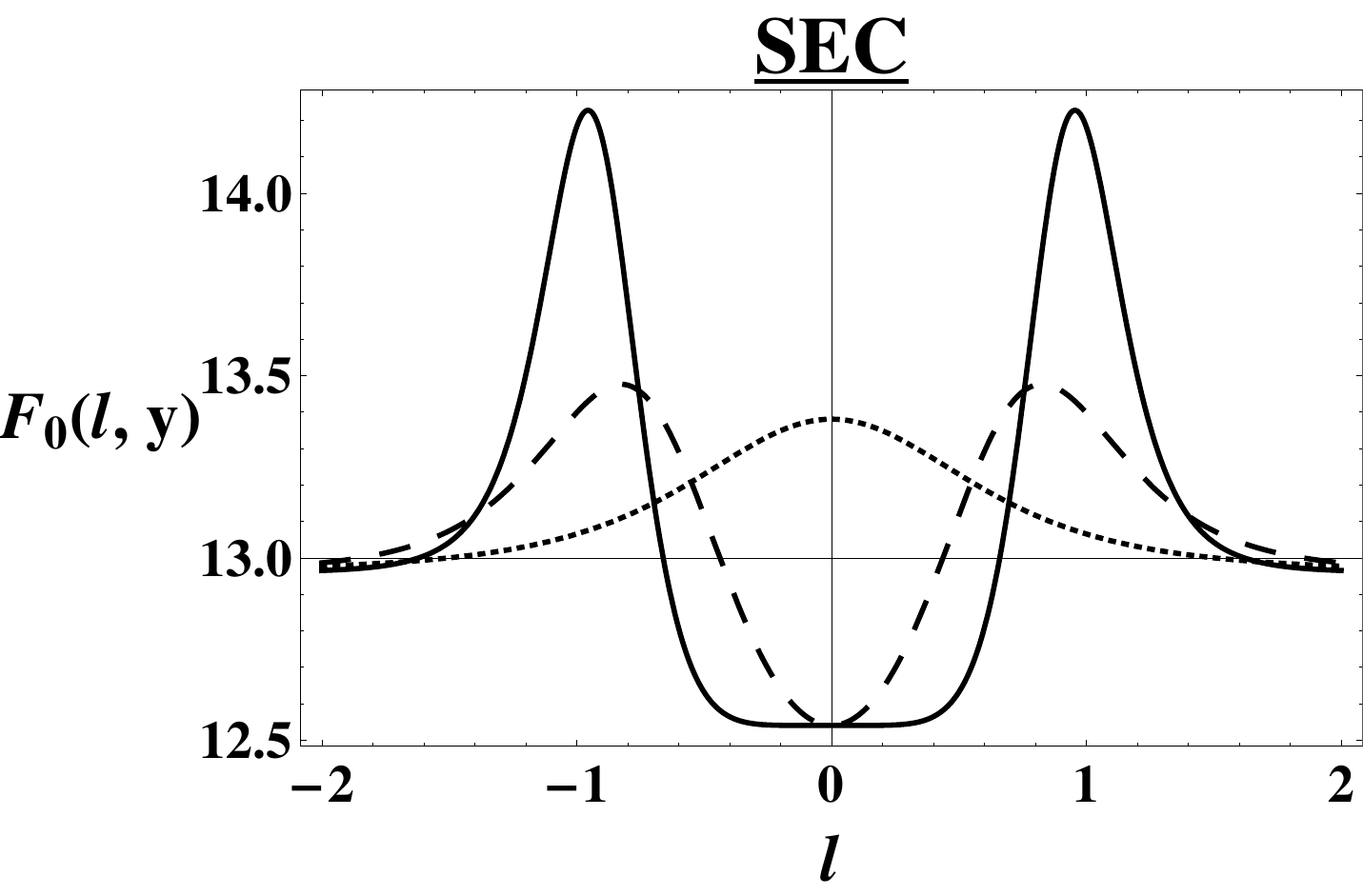} 
\caption{Profile of $F_0^{(S)}$ of for $f(y) = -\log[\cosh(y)]$ (left plot) and $f(y) = \log[\cosh(y)]$ (right plot) at $|y| = 1$ with $b_0 = 1$ and $m=2$ (dotted curve), $m = 4$ (dashed curve) and $m = 8$ (continuous curve).} \la{fig:SEC-f0}
\end{figure}



\subsection{Inequalities of DEC}

The inequality functions for DEC in our 5D model is given below. It is clear that the DEC would not satisfy for a growing warp factor. However, for the decaying warp factor, these conditions may be satisfied in limited domain on the $l-y$ space.   
\begin{eqnarray}
F_{1}^{(D)} \le(l, y \ri) &=& \rho(l, y) - \lvert \tau(l, y) \rvert \nonumber \\
&=& - \Big \lvert \frac{e^{-2f(y)} \Big( -1 + r'(l)^{2} \Big)}{r(l)^{2}} + 3 \Big( 2f'(y)^{2} + f''(y) \Big) \Big \rvert -3 \Big( 2f'(y)^{2} + f''(y) \Big)  \nonumber \\
&&~~~~~~~~~~~~~~ - \frac{e^{-2f(y)} \Big( -1 + r'(l)^{2} + 2r(l)r''(l) \Big)}{r(l)^{2}} ,
\label{eq:DEC-F0}
\end{eqnarray}
\begin{eqnarray}
F_{2}^{(D)} \le(l, y \ri) &=& \rho(l, y) - \lvert p_{2}(l, y) \rvert = F_{3}^{(D)}(l, y)  \nonumber \\
&=& - \Big \lvert 6f'(y)^{2} + 3f''(y) + \frac{e^{-2f(y)}r''(l)}{r(l)} \Big \rvert -3 \Big( 2f'(y)^{2} + f''(y) \Big) \nonumber \\
&&~~~~~~~~ - \frac{e^{-2f(y)} \Big( -1 + r'(l)^{2} + 2r(l)r''(l) \Big)}{r(l)^{2}} ,
\label{eq:DEC-F1}
\end{eqnarray}
\begin{eqnarray}
F_{4}^{(D)}(l, y) &=& \rho(l, y) - \lvert p_{4}(l, y) \rvert \nonumber \\
&=& - \Big \lvert 6f'(y)^{2} + \frac{e^{-2f(y) \big( -1 + r'(l)^{2} + 2r(l)r''(l) \big)}}{r(l)^{2}} \Big \rvert - 3\Big( 2f'(y)^{2} + f''(y) \Big) \nonumber \\
&&~~~~~~~~~~~~- \frac{e^{-2f(y)} \Big( -1 + r'(l)^{2} + 2r(l)r''(l) \Big)}{r(l)^{2}} .
\label{eq:DEC-F3}
\end{eqnarray}
The functional dependence of the inequality functions on $l$ at the location of the brane, for decaying warp factor, is written down in Table \ref{tab:WDEC}. 

\begin{center}
\begin{adjustbox}{max width=\textwidth}
\begin{tabular}{|c|c|c|c|}
\hline
Ineq. Fns. &$m=2$& $m = 4$ & $m = 8$ \\ \hline
$F_{1}^{(D)}$ & $3 - \frac{1}{(1+l^{2})^{2}} - \Big \lvert 3 + \frac{1}{(1+l^{2})^{2}} \Big \rvert$ & $3 + \frac{1}{\sqrt{1+l^{4}}} - \frac{l^{2}(6+l^{4})}{(1+l^{4})^{2}} - \Big \lvert 3 - \frac{l^{6}}{(1+l^{4})^{2}} + \frac{1}{\sqrt{1+l^{4}}} \Big \rvert $ & $3 + \frac{1}{(1+l^{8})^{1/4}} - \frac{l^{6}(14+l^{8})}{(1+l^{8})^{2}} - \Big \lvert 3 - \frac{l^{14}}{(1+l^{8})^{2}} + \frac{1}{(1+l^{8})^{1/4}} \Big \rvert $ \\ \hline
$F_{2}^{(D)} = F_{3}^{(D)}$ & $3 - \frac{1}{(1+l^{2})^{2}} - \Big \lvert 3 - \frac{1}{(1+l^{2})^{2}} \Big \rvert$ & $3 + \frac{1}{\sqrt{1+l^{4}}} - \frac{l^{2}(6+l^{4})}{(1+l^{4})^{2}} -3 \Big \lvert 1 - \frac{l^{2}}{(1+l^{4})^{2}} \Big \rvert $ & $3 + \frac{1}{(1+l^{8})^{1/4}} - \frac{l^{6}(14+l^{8})}{(1+l^{8})^{2}} - \Big \lvert 3 - \frac{7l^{6}}{(1+l^{8})^{2}} \Big \rvert $ \\ \hline
$F_{4}^{(D)}$ & $3 - \frac{1}{(1+l^{2})^{2}} - \Big \lvert \frac{1}{(1+l^{2})^{2}} \Big \rvert$ & $3 + \frac{1}{\sqrt{1+l^{4}}} - \frac{l^{2}(6+l^{4})}{(1+l^{4})^{2}} - \Big \lvert \frac{1}{\sqrt{1+l^{4}}} - \frac{l^{2}(6+l^{4})}{(1+l^{4})^{2}}\Big \rvert $ & $3 + \frac{1}{(1+l^{8})^{1/4}} - \frac{l^{6}(14+l^{8})}{(1+l^{8})^{2}} - \Big \lvert \frac{1}{(1+l^{8})^{1/4}} - \frac{l^{6}(14+l^{8})}{(1+l^{8})^{2}} \Big \rvert $ \\ \hline
\end{tabular}
\end{adjustbox}
\captionof{table}{Inequality functions of DEC for decaying warp factor, at $y = 0$, $b_{0} = 1$, $m =2, 4$ and $8$.} \label{tab:WDEC}
\end{center}

Note that, at $y=0$, for $m=2$, $F_{1}^{(D)}$ is always negative whereas $F_{2}^{(D)}$ (and $F_{3}^{(D)}$) identically vanishes everywhere and $F_{4}^{(D)}=3$ in $l$. 
In case of $m = 4$ (or $m=8$), the sum of the first two terms of $F_{1}^{(D)}$ will always be less than the sum of the last two terms for all values of $l$ (except for $l = 0$), thus the first inequality of DEC is always violated except at the wormhole throat even for the decaying warp factor. 
However, $F_{2}^{(D)}$ (and $F_{3}^{(D)}$) is positive for restricted domain of coordinate $l$ $\approx $($-0.62 \leq l \leq 0.62$ for $m=4$ and $-0.73 \leq l \leq 0.73$ for $m=8$. 
Whereas $F_{4}^{(D)}$ is constant, for both $m = 4$ and $m = 8$, in the domain  ($-0.41 \lesssim l \lesssim 0.41$) and ($-0.64 \lesssim l \lesssim 0.64$) respectively but beyond these domain of $l$, $F_{4}^{(D)}$ can be negative or positive. In general one can say that the fourth inequality of DEC is always satisfied for $m=4$ and partially satisfied for $m=8$.

We plot a limited part of the parameter space in $l-y$ plane where the second (Fig. \ref{fig:5DEC-f2}) and fourth (Fig. \ref{fig:5DEC-f4}) inequalities of DEC are satisfied. Fig. \ref{fig:5DEC-f2} suggests that in the limit $m\ra \infty$ and $y \ra 0$, the domain of $l$ where the inequality is satisfied increases. On the other hand, Fig. \ref{fig:5DEC-f4} implies that, in the limit $m\ra \infty$, the inequality is satisfied only at $y \ra 0$ and $l\ra 0$, i.e. on the brane and near the throat.
\begin{figure}[H]
\centering
 \includegraphics[scale=.25]{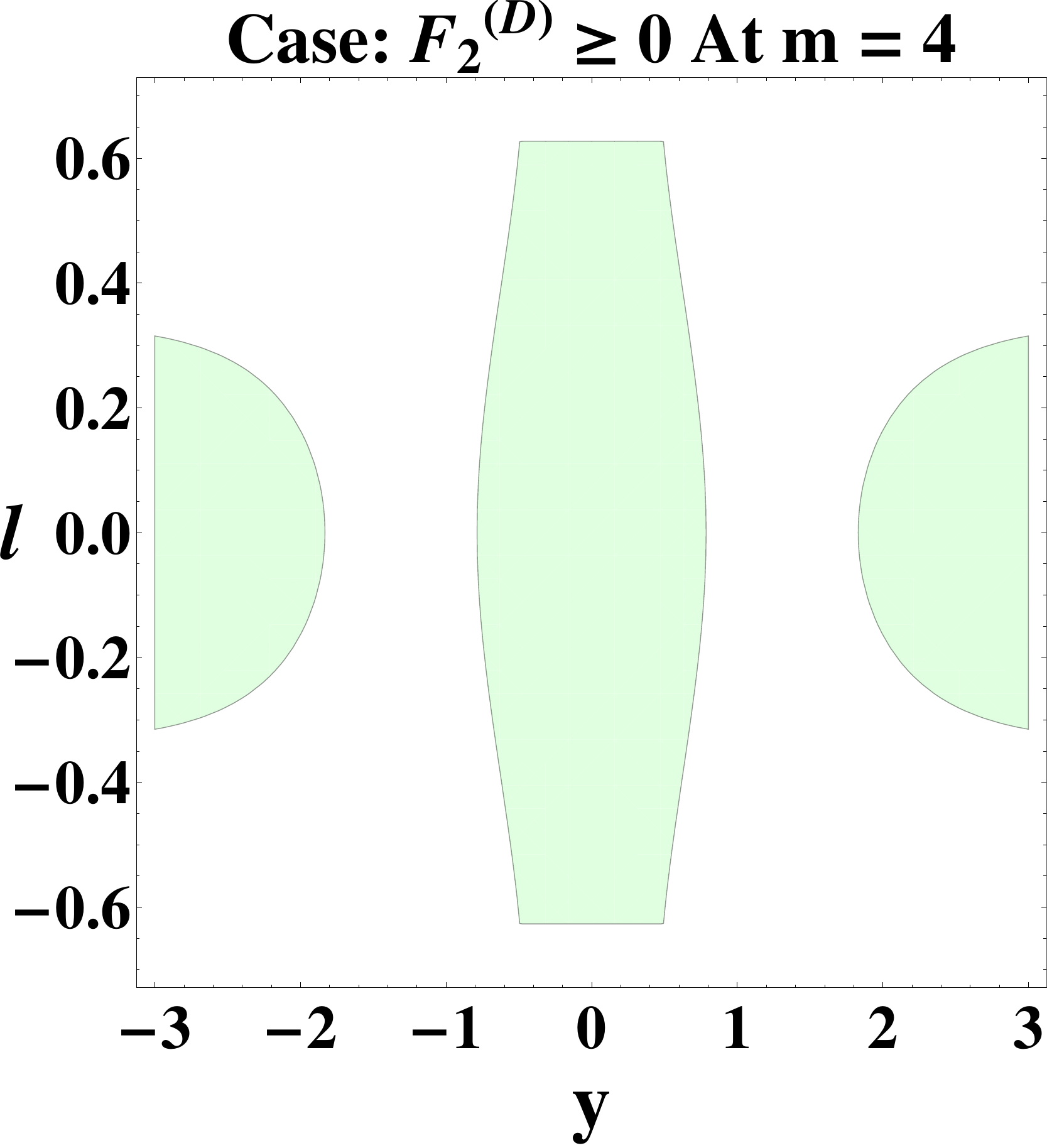}
 \hspace{1cm}
 \includegraphics[scale=.25]{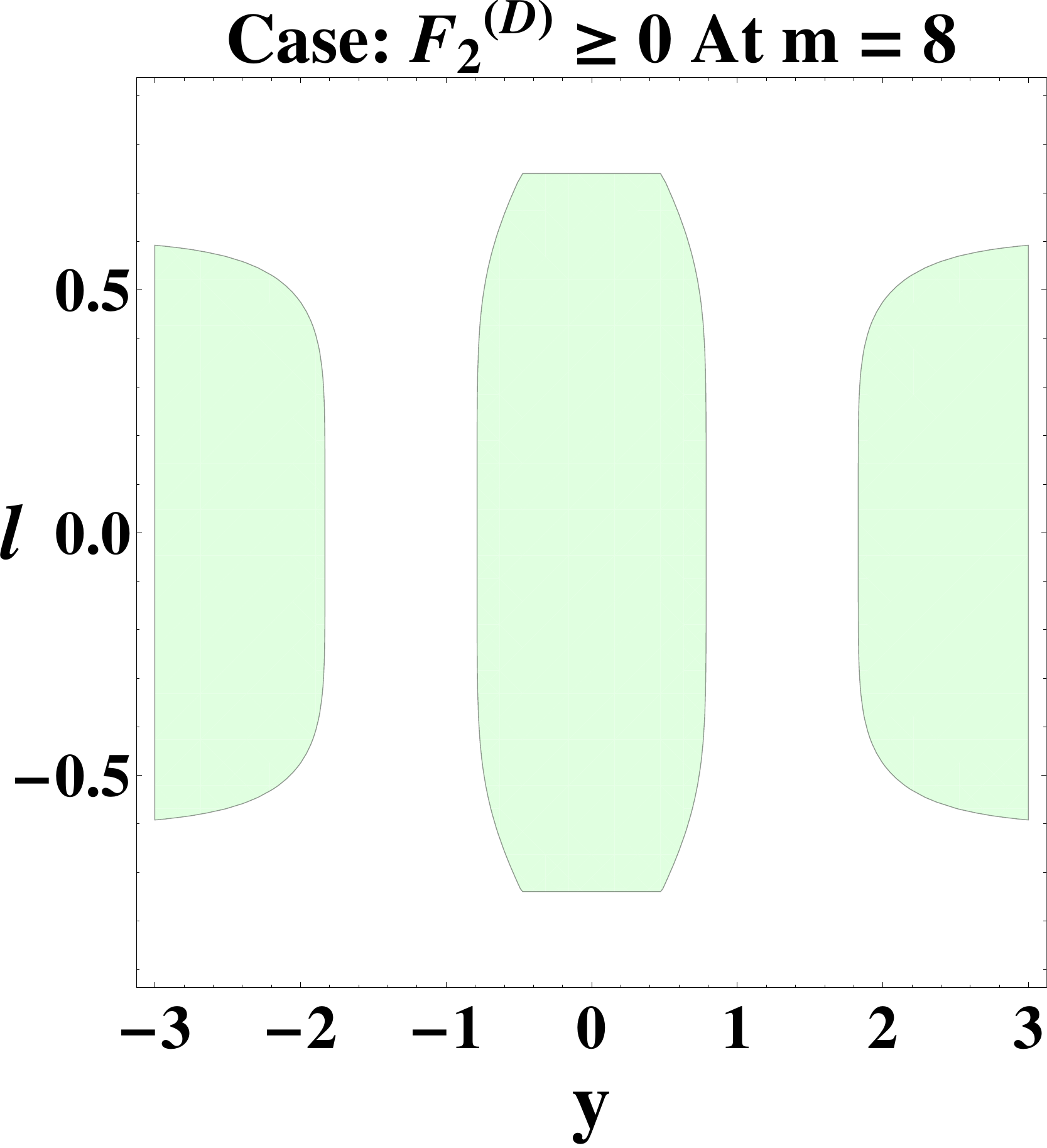} 
 \caption{Parameter space plots $F_{2}^{(D)}>0$ (where $f(y) = -\log[\cosh(y)]$) for $m = 4$ and $8$.} \la{fig:5DEC-f2}
\end{figure}
\begin{figure}[H]
\centering
 \includegraphics[scale=.25]{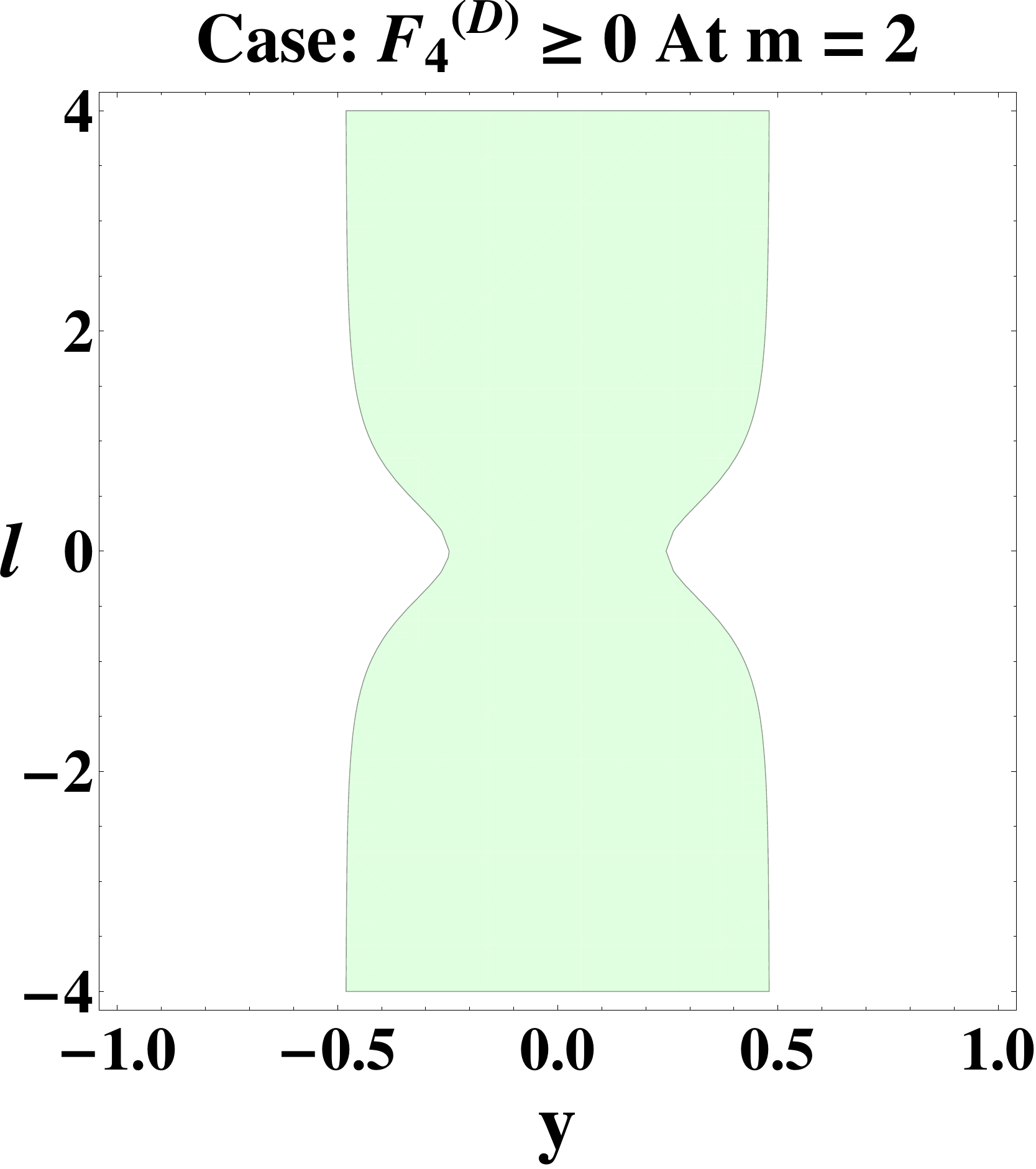}
 \hspace{1cm}
 \includegraphics[scale=.25]{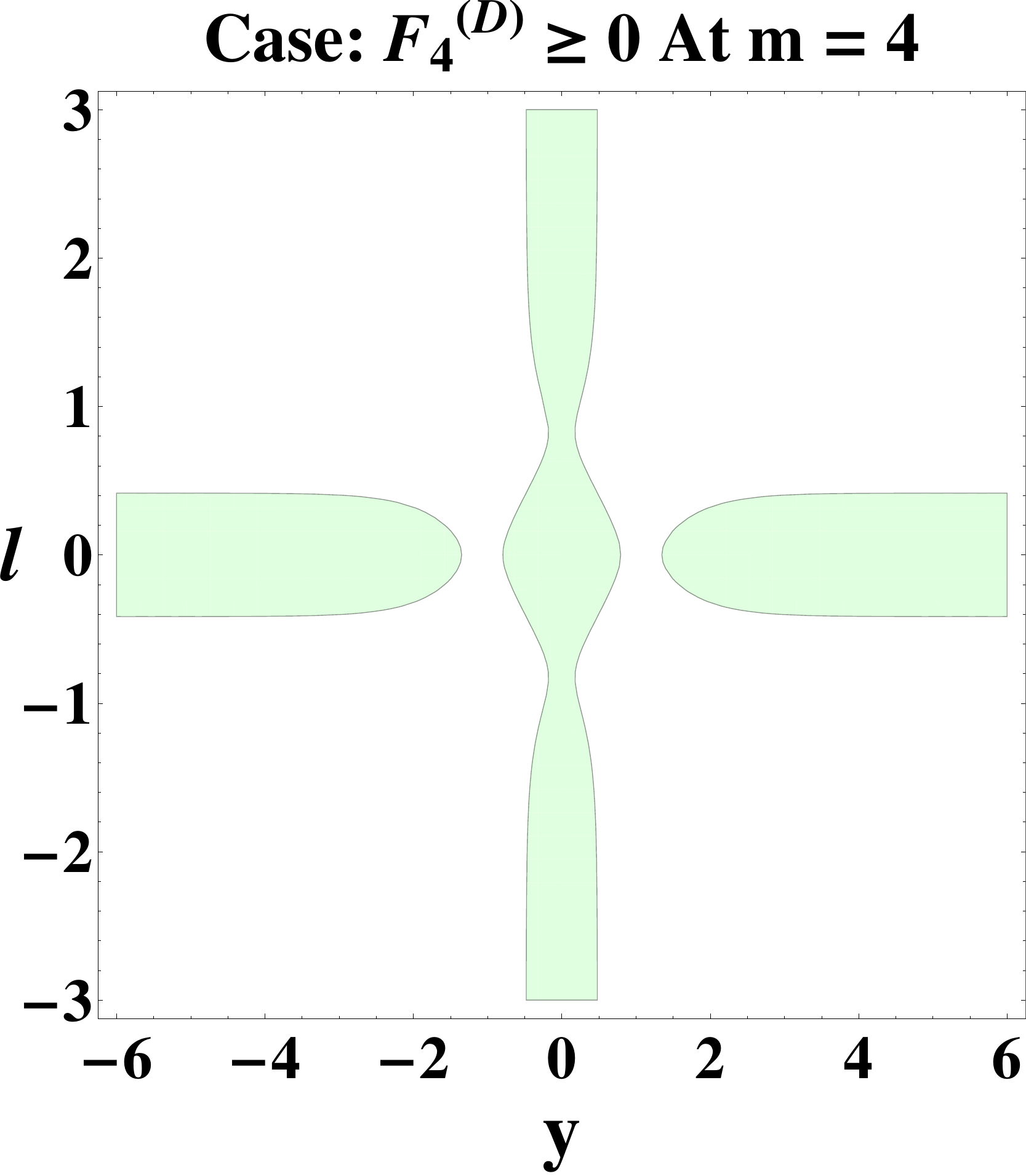}
 \hspace{1cm}
 \includegraphics[scale=.25]{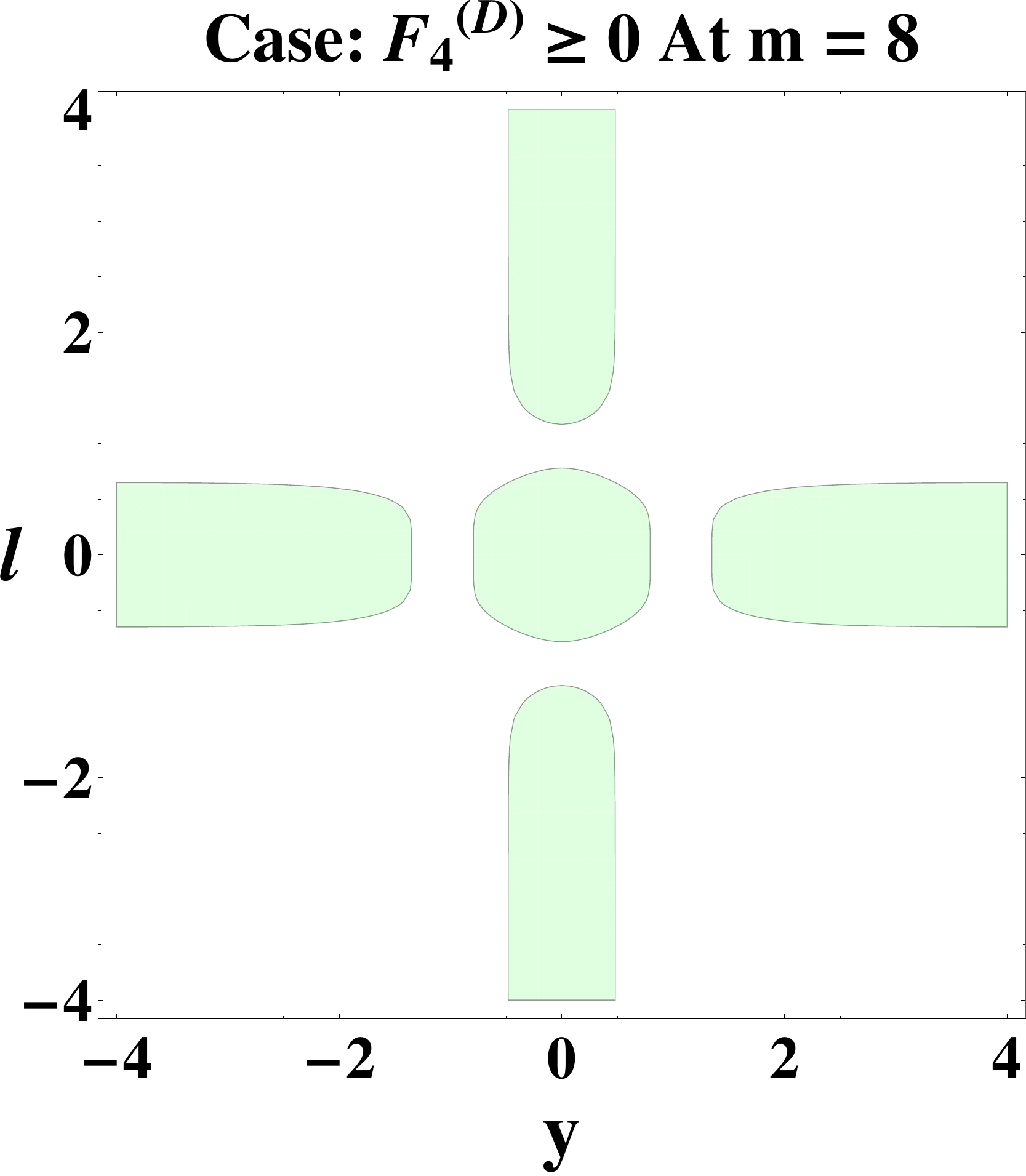} 
 \caption{Parameter space plots $F_{4}^{(D)}>0$ (where $f(y) = -\log[\cosh(y)]$) for $m =2, 4$ and $8$.} \la{fig:5DEC-f4}
\end{figure}
After discussing the energy conditions for a GEB wormhole embedded in a 5D warped braneworld background in detail, we summarise in Table \ref{tab:EC-WGEB} \footnote{Here the status in presence decaying and growing warp factor is separated by `/'}, the status of the inequalities of the energy conditions, at or near the location of the so-called brane. Note that energy conditions away from $y=0$ have also been analysed and it has been shown that they indeed are satisfied in few cases particularly in presence of decaying warp factor. We discuss the results in the next section.

\begin{center}

\begin{adjustbox}{max width=\textwidth}
\begin{tabular}{|l|*{9}{c|}}
\hline
\multicolumn{1}{|l|}{Warped GEB} & \multicolumn{1}{|l|}{} &   \multicolumn{3}{|c|}{$m=2$} & \multicolumn{1}{|l|}{}   & \multicolumn{3}{|c|}{$m>2$}\\ \hline

Ineq. Fns. & &   WEC   &   SEC   & DEC               &    &  WEC   &   SEC   & DEC \\ \hline  \hline

$F_{0} \geq 0$ & & Yes / No & No / Yes & --      &    & Yes / No & No / Yes & --  \\ \hline

$F_{1} \geq 0$ & & No / No & No / No & No / No &    & No / No & No / No & No /No \\ \hline

$F_{2},F_{3} \geq 0$ & & Yes / Yes & Y / Y & Yes / No    &    & Y / Y & Y / Y & Y / No \\ \hline


$F_{4} \geq 0$ & & Yes / No  & Yes / No & Y / No &   & Yes / No  & Yes / No & Y / No  \\ \hline

\end{tabular}

\end{adjustbox}

\captionof{table}{Status of the inequalities of the energy conditions for our 5D wormhole-geometry, with "decaying / growing warp factor", at $y \ra 0 $ where `Yes', `Y' and `No' stands for (as before) satisfied, partially satisfied and not satisfied everywhere in $l$ respectively.}
\label{tab:EC-WGEB}
\end{center}

\section{Discussion}

Existence of wormholes and extra dimensions, though naturally appears in theoretical physics as solutions of Einstein equations and in the context of unified field theories are yet to be confirmed through observations/experiments.  Wormhole solutions, within the context of general theory of relativity, has serious drawbacks as they do not satisfy energy conditions and need negative energy density to support the wormhole `throat'. On the other hand, extra dimensions though have not been discovered, required by fundamental symmetries and provide new possibilities. 
In this article, we provide a new way to avoid exotic matter for wormhole geometry as well as we demonstrate yet another advantage of having a warped extra spatial dimension by embedding a GEB wormhole in a five dimensional warped braneworld background.
Comparison between Table \ref{tab:EC-GEB} and Table \ref{tab:EC-WGEB} summarises the key results and we discuss them in a systematic manner in the following. 
\begin{itemize}
\item Table \ref{tab:EC-GEB} shows that, in the 4D scenario, it is the generalisation of the Ellis-Bronnikov model and setting the new parameter $m > 2$  ($m=2$ case corresponds to the original EB model) that provides WEC satisfying wormhole geometry. However the WEC is satisfied near throat and violated near $l\ra \pm b_0$. With increasing $m$ the magnitude of violation increases though the negative density matter accumulates in ever-narrower region near $l\ra \pm b_0$. 
\item The location $l=b_0$ is a challenging barrier for a positive energy particle to cross without interacting with matter of negative energy density. However, as the negative energy spike goes deeper with increasing $m$, `stronger' becomes the throat as suggested by the increased flatness of $f_0^{(W)}$ around $l=0$ (upto $m$-th derivative of $f_0^{(W)}$ vanishes leading to increased stability of the structure with increasing $m$ against small perturbations).

\item On the other hand, Table \ref{tab:EC-WGEB} shows, {\em presence of a warped extra dimension (with a decaying warp factor) leads to  satisfaction of WEC and renders the energy density positive even in the case of $m=2$}. Further for $m>2$, status of inequalities have improved significantly compared to the 4D GEB model.
One of the key aspects of the 5D model is to resolve the particular drawback mentioned above.
\item Physical understanding behind the rise of positive energy density is the following. In thick brane models, it has been shown that \cite{Dzhunushaliev:2009va}, a warped geometry with growing warp factor is sourced by bulk phantom or tachyon fields, whereas a decaying warp factor shows up in presence of matter fields with positive energy density, e.g. a scalar field with Sine-Gordon potential. Thus in presence of a decaying warp factor, the negative energy density (needed in 4D EB scenario) is compensated for. 
\end{itemize}

A complete picture of the role played by the warped extra dimension, particularly the one with a decaying warp factor, towards forming a viable wormhole geometry can be realised by looking at the geodetic potentials, particle trajectories and evolution of geodesic congruences as such. One may also ask, to what extent a EB wormhole embedded in 5D braneworld with a decaying warp factor can mimic physical aspects GEB model. These and studies on other astrophysical aspects such as lensing effect, stability of the wormhole will be reported in future communications.

\section*{Acknowledgments}

This research is supported by a start-up grant awarded by the University Grant Commission, Govt. of India, with grant number no. F.30-420/2018(BSR). The authors would like to thank Sayan Kar and Saibal Ray for helpful discussions and suggestions.
   
\section*{Bibliography} 


\begin{thebibliography}{99}
%

\bibitem{Visser-book}
M. Visser, Lorentzian wormholes: from Einstein to Hawking, AIP Press (New York) 1995.


\bibitem{Hawking-book} S. W. Hawking and G. F. R. Ellis, The large scale structure of spacetime, Cambridge University Press, 1973.

\bibitem{Wald-book}
R. M. Wald, General Relativity, University of Chicago Press, First Indian Edition (2006).

\bibitem{Lobo-book}
F.S.N. Lobo (ed.), Wormholes, Warp Drives and Energy Conditions, Springer (2017).


\bibitem{Witten:2019qhl}
E.~Witten,
``Light Rays, Singularities, and All That,''
Rev. Mod. Phys. \textbf{92}, no.4, 045004 (2020)
[arXiv:1901.03928 [hep-th]].


\bibitem{Flamm-1916}
L. Flamm, ``Comments on Einstein’s Theory of Gravity,” Physikalische Zeitschrift {\bf 17}, 448 (1916).


\bibitem{Einstein:1935tc}
A.~Einstein and N.~Rosen,
``The Particle Problem in the General Theory of Relativity,''
Phys. Rev. \textbf{48}, 73-77 (1935).

\bibitem{Fuller:1962zza}
R.~W.~Fuller and J.~A.~Wheeler,
``Causality and Multiply Connected Space-Time,''
Phys. Rev. \textbf{128}, 919-929 (1962)

\bibitem{Morris:1988tu}
M.~S.~Morris, K.~S.~Thorne and U.~Yurtsever,
``Wormholes, Time Machines, and the Weak Energy Condition,''
Phys. Rev. Lett. \textbf{61}, 1446-1449 (1988).

\bibitem{Morris:1988cz}
M.~S.~Morris and K.~S.~Thorne,
``Wormholes in space-time and their use for interstellar travel: A tool for teaching general relativity,''
Am. J. Phys. \textbf{56}, 395-412 (1988).

\bibitem{Lobo:2017oab}
M.~Alcubierre and F.~S.~N.~Lobo,
``Wormholes, Warp Drives and Energy Conditions,''
Fundam. Theor. Phys. \textbf{189}, pp.-279 (2017).
[arXiv:2103.05610 [gr-qc]].

\bibitem{Roman:2004xm}
T.~A.~Roman,
``Some thoughts on energy conditions and wormholes,''
[arXiv:gr-qc/0409090 [gr-qc]].


\bibitem{Hochberg:1998ha}
D.~Hochberg and M.~Visser,
``Dynamic wormholes, anti-trapped surfaces, and energy conditions,''
Phys. Rev. D \textbf{58}, 044021 (1998)
[arXiv:gr-qc/9802046 [gr-qc]].


\bibitem{Gao:2016bin}
P.~Gao, D.~L.~Jafferis and A.~C.~Wall,
``Traversable Wormholes via a Double Trace Deformation,''
JHEP \textbf{12}, 151 (2017)
[arXiv:1608.05687 [hep-th]].

\bibitem{Maldacena:2018gjk}
J.~Maldacena, A.~Milekhin and F.~Popov,
``Traversable wormholes in four dimensions,''
[arXiv:1807.04726 [hep-th]].


\bibitem{Fu:2018oaq}
Z.~Fu, B.~Grado-White and D.~Marolf,
``A perturbative perspective on self-supporting wormholes,''
Class. Quant. Grav. \textbf{36}, no.4, 045006 (2019)
[erratum: Class. Quant. Grav. \textbf{36}, no.24, 249501 (2019)]
[arXiv:1807.07917 [hep-th]].


\bibitem{Fu:2019vco}
Z.~Fu, B.~Grado-White and D.~Marolf,
``Traversable Asymptotically Flat Wormholes with Short Transit Times,''
Class. Quant. Grav. \textbf{36}, no.24, 245018 (2019)
[arXiv:1908.03273 [hep-th]].



%

%




\bibitem{Hochberg:1990is}
D.~Hochberg,
``Lorentzian wormholes in higher order gravity theories,''
Phys. Lett. B \textbf{251}, 349-354 (1990).

\bibitem{Bhawal:1992sz}
B.~Bhawal and S.~Kar,
``Lorentzian wormholes in Einstein-Gauss-Bonnet theory,''
Phys. Rev. D \textbf{46}, 2464-2468 (1992).

\bibitem{Agnese:1995kd}
A.~G.~Agnese and M.~La Camera,
``Wormholes in the Brans-Dicke theory of gravitation,''
Phys. Rev. D \textbf{51}, 2011-2013 (1995).

\bibitem{Samanta:2018hbw}
G.~C.~Samanta, N.~Godani and K.~Bamba,
``Traversable wormholes with exponential shape function in modified gravity and general relativity: A comparative study,''
Int. J. Mod. Phys. D \textbf{29}, no.09, 2050068 (2020)
[arXiv:1811.06834 [gr-qc]].



\bibitem{Lobo:2008zu}
F.~S.~N.~Lobo,
``General class of wormhole geometries in conformal Weyl gravity,''
Class. Quant. Grav. \textbf{25}, 175006 (2008)
[arXiv:0801.4401 [gr-qc]].


\bibitem{Kanti:2011jz}
P.~Kanti, B.~Kleihaus and J.~Kunz,
``Wormholes in Dilatonic Einstein-Gauss-Bonnet Theory,''
Phys. Rev. Lett. \textbf{107}, 271101 (2011)
[arXiv:1108.3003 [gr-qc]].

\bibitem{Kanti:2011yv}
P.~Kanti, B.~Kleihaus and J.~Kunz,
``Stable Lorentzian Wormholes in Dilatonic Einstein-Gauss-Bonnet Theory,''
Phys. Rev. D \textbf{85}, 044007 (2012)
[arXiv:1111.4049 [hep-th]].


\bibitem{Zubair:2017oir}
M.~Zubair, F.~Kousar and S.~Bahamonde,
``Static spherically symmetric wormholes in generalized $f(R,\phi)$ gravity,''
Eur. Phys. J. Plus \textbf{133}, no.12, 523 (2018)
[arXiv:1712.05699 [gr-qc]].


\bibitem{Shaikh:2016dpl}
R.~Shaikh and S.~Kar,
``Wormholes, the weak energy condition, and scalar-tensor gravity,''
Phys. Rev. D \textbf{94}, no.2, 024011 (2016)
[arXiv:1604.02857 [gr-qc]].


\bibitem{Ovgun:2018xys}
A.~\"Ovg\"un, K.~Jusufi and \.I.~Sakall\i{},
``Exact traversable wormhole solution in bumblebee gravity,''
Phys. Rev. D \textbf{99}, no.2, 024042 (2019)
[arXiv:1804.09911 [gr-qc]].


\bibitem{Canate:2019spb}
P.~Ca\~nate, J.~Sultana and D.~Kazanas,
``Ellis wormhole without a phantom scalar field,''
Phys. Rev. D \textbf{100}, no.6, 064007 (2019)
[arXiv:1907.09463 [gr-qc]].





\bibitem{Hochberg:1998ii}
D.~Hochberg and M.~Visser,
``The Null energy condition in dynamic wormholes,''
Phys. Rev. Lett. \textbf{81}, 746-749 (1998)
[arXiv:gr-qc/9802048 [gr-qc]].

\bibitem{Roman:1992xj}
T.~A.~Roman,
``Inflating Lorentzian wormholes,''
Phys. Rev. D \textbf{47}, 1370-1379 (1993)
[arXiv:gr-qc/9211012 [gr-qc]].

\bibitem{Kar:1994tz}
S.~Kar,
``Evolving wormholes and the weak energy condition,''
Phys. Rev. D \textbf{49}, 862-865 (1994).

\bibitem{Kar:1995ss}
S.~Kar and D.~Sahdev,
``Evolving Lorentzian wormholes,''
Phys. Rev. D \textbf{53}, 722-730 (1996)
[arXiv:gr-qc/9506094 [gr-qc]].


\bibitem{Visser:2003yf}
M.~Visser, S.~Kar and N.~Dadhich,
``Traversable wormholes with arbitrarily small energy condition violations,''
Phys. Rev. Lett. \textbf{90}, 201102 (2003)
[arXiv:gr-qc/0301003 [gr-qc]].






\bibitem{Lobo:2009ip}
F.~S.~N.~Lobo and M.~A.~Oliveira,
``Wormhole geometries in f(R) modified theories of gravity,''
Phys. Rev. D \textbf{80}, 104012 (2009)
[arXiv:0909.5539 [gr-qc]].


\bibitem{Garcia:2010xb}
N.~M.~Garcia and F.~S.~N.~Lobo,
``Wormhole geometries supported by a nonminimal curvature-matter coupling,''
Phys. Rev. D \textbf{82}, 104018 (2010)
[arXiv:1007.3040 [gr-qc]].

\bibitem{MontelongoGarcia:2010xd}
N.~Montelongo Garcia and F.~S.~N.~Lobo,
``Nonminimal curvature-matter coupled wormholes with matter satisfying the null energy condition,''
Class. Quant. Grav. \textbf{28}, 085018 (2011)
[arXiv:1012.2443 [gr-qc]].


\bibitem{Sajadi:2012zz}
S.~N.~Sajadi and N.~Riazi,
``Expanding lorentzian wormholes in R**2 gravity,''
Prog. Theor. Phys. \textbf{126}, 753-760 (2011)
doi:10.1143/PTP.126.753.






\bibitem{Bertolami:2012fz}
O.~Bertolami and R.~Zambujal Ferreira,
``Traversable Wormholes and Time Machines in non-minimally coupled curvature-matter $f(R)$ theories,''
Phys. Rev. D \textbf{85}, 104050 (2012)
[arXiv:1203.0523 [gr-qc]].

\bibitem{Harko:2013yb}
T.~Harko, F.~S.~N.~Lobo, M.~K.~Mak and S.~V.~Sushkov,
``Modified-gravity wormholes without exotic matter,''
Phys. Rev. D \textbf{87}, no.6, 067504 (2013)
[arXiv:1301.6878 [gr-qc]].

\bibitem{Pavlovic:2014gba}
P.~Pavlovic and M.~Sossich,
``Wormholes in viable $f(R)$ modified theories of gravity and Weak Energy Condition,''
Eur. Phys. J. C \textbf{75}, 117 (2015)
[arXiv:1406.2509 [gr-qc]].

\bibitem{Varieschi:2015wwa}
G.~U.~Varieschi and K.~L.~Ault,
``Wormhole geometries in fourth-order conformal Weyl gravity,''
Int. J. Mod. Phys. D \textbf{25}, no.06, 1650064 (2016)
[arXiv:1510.05054 [gr-qc]].

\bibitem{Zangeneh:2015jda}
M.~Kord Zangeneh, F.~S.~N.~Lobo and M.~H.~Dehghani,
``Traversable wormholes satisfying the weak energy condition in third-order Lovelock gravity,''
Phys. Rev. D \textbf{92}, no.12, 124049 (2015)
[arXiv:1510.07089 [gr-qc]].

\bibitem{Duplessis:2015xva}
F.~Duplessis and D.~A.~Easson,
``Traversable wormholes and non-singular black holes from the vacuum of quadratic gravity,''
Phys. Rev. D \textbf{92}, no.4, 043516 (2015)
[arXiv:1506.00988 [gr-qc]].


\bibitem{Mehdizadeh:2015jra}
M.~R.~Mehdizadeh, M.~Kord Zangeneh and F.~S.~N.~Lobo,
``Einstein-Gauss-Bonnet traversable wormholes satisfying the weak energy condition,''
Phys. Rev. D \textbf{91}, no.8, 084004 (2015)
[arXiv:1501.04773 [gr-qc]].

\bibitem{Samanta:2019tjb}
G.~C.~Samanta and N.~Godani,
``Validation of energy conditions in wormhole geometry within viable $f(R)$ gravity,''
Eur. Phys. J. C \textbf{79}, no.7, 623 (2019)
[arXiv:1908.04406 [gr-qc]].

\bibitem{Godani:2018blx}
N.~Godani and G.~C.~Samanta,
``Traversable Wormholes and Energy Conditions with Two Different Shape Functions in $f(R)$ Gravity,''
Int. J. Mod. Phys. D \textbf{28}, no.02, 1950039 (2018)
[arXiv:1809.00341 [gr-qc]].



\bibitem{Rahaman:2014dpa}
F.~Rahaman, I.~Karar, S.~Karmakar and S.~Ray,
``Wormhole inspired by non-commutative geometry,''
Phys. Lett. B \textbf{746} (2015), 73-78
[arXiv:1406.3045 [gr-qc]].



\bibitem{Rahaman:2016jds}
F.~Rahaman, N.~Paul, A.~Banerjee, S.~S.~De, S.~Ray and A.~A.~Usmani,
``The Finslerian wormhole models,''
Eur. Phys. J. C \textbf{76} (2016) no.5, 246
[arXiv:1607.04329 [gr-qc]].


\bibitem{Chakraborty:2021qwl}
K.~Chakraborty, A.~Aziz, F.~Rahaman and S.~Ray,
``Quark matter supported wormhole in third order Lovelock gravity,''
[arXiv:2104.13966 [gr-qc]].

\bibitem{Mishra:2021ato}
B.~Mishra, A.~S.~Agrawal, S.~K.~Tripathy and S.~K.~T.~S.~Ray,
``Wormhole solutions in f(R) gravity,''
Int. J. Mod. Phys. D \textbf{30} (2021) no.08, 2150061
[arXiv:2104.05440 [gr-qc]].





\bibitem{Maeda:2008nz}
H.~Maeda and M.~Nozawa,
``Static and symmetric wormholes respecting energy conditions in Einstein-Gauss-Bonnet gravity,''
Phys. Rev. D \textbf{78}, 024005 (2008)
[arXiv:0803.1704 [gr-qc]].

\bibitem{Shaikh:2015oha}
R.~Shaikh,
``Lorentzian wormholes in Eddington-inspired Born-Infeld gravity,''
Phys. Rev. D \textbf{92}, 024015 (2015)
[arXiv:1505.01314 [gr-qc]].


\bibitem{Shaikh:2018yku}
R.~Shaikh,
``Wormholes with nonexotic matter in Born-Infeld gravity,''
Phys. Rev. D \textbf{98}, no.6, 064033 (2018)
[arXiv:1807.07941 [gr-qc]].




\bibitem{Bohmer:2011si}
C.~G.~Bo{\"o}hmer, T.~Harko and F.~S.~N.~Lobo,
``Wormhole geometries in modified teleparralel gravity and the energy conditions,''
Phys. Rev. D \textbf{85}, 044033 (2012)
[arXiv:1110.5756 [gr-qc]].


\bibitem{Bronnikov:2015pha}
K.~A.~Bronnikov and A.~M.~Galiakhmetov,
``Wormholes without exotic matter in Einstein\textendash{}Cartan theory,''
Grav. Cosmol. \textbf{21}, no.4, 283-288 (2015)
[arXiv:1508.01114 [gr-qc]].

\bibitem{Battista:2017hyj}
E.~Di Grezia, E.~Battista, M.~Manfredonia and G.~Miele,
``Spin, torsion and violation of null energy condition in traversable wormholes,''
Eur. Phys. J. Plus \textbf{132}, no.12, 537 (2017)
[arXiv:1707.01508 [gr-qc]].







\bibitem{Cramer:1994qj}
J.~G.~Cramer, R.~L.~Forward, M.~S.~Morris, M.~Visser, G.~Benford and G.~A.~Landis,
``Natural wormholes as gravitational lenses,''
Phys. Rev. D \textbf{51}, 3117-3120 (1995)
[arXiv:astro-ph/9409051].

\bibitem{Dey:2008kn}
T.~K.~Dey and S.~Sen,
``Gravitational lensing by wormholes,''
Mod. Phys. Lett. A \textbf{23}, 953-962 (2008)
[arXiv:0806.4059 [gr-qc]].

\bibitem{Shaikh:2018oul}
R.~Shaikh, P.~Banerjee, S.~Paul and T.~Sarkar,
``A novel gravitational lensing feature by wormholes,''
Phys. Lett. B \textbf{789}, 270-275 (2019)
[erratum: Phys. Lett. B \textbf{791}, 422-423 (2019)]
[arXiv:1811.08245 [gr-qc]].


\bibitem{Bishop:2021rye}
N.~T.~Bishop,
``Introduction to gravitational wave astronomy,''
[arXiv:2103.07675 [gr-qc]].

\bibitem{Arimoto:2021cwc}
M.~Arimoto
\textit{et al.}
``Gravitational Wave Physics and Astronomy in the nascent era,''
[arXiv:2104.02445 [gr-qc]].

\bibitem{Dent:2020nfa}
J.~B.~Dent, W.~E.~Gabella, K.~Holley-Bockelmann and T.~W.~Kephart,
``The Sound of Clearing the Throat: Gravitational Waves from a Black Hole Orbiting in a Wormhole Geometry,''
[arXiv:2007.09135 [gr-qc]].

\bibitem{Akiyama:2019cqa}
K.~Akiyama \textit{et al.} [Event Horizon Telescope],
``First M87 Event Horizon Telescope Results. I. The Shadow of the Supermassive Black Hole,''
Astrophys. J. Lett. \textbf{875}, L1 (2019)
[arXiv:1906.11238 [astro-ph.GA]].


\bibitem{Rahaman:2014pba}
F.~Rahaman, P.~Salucci, P.~K.~F.~Kuhfittig, S.~Ray and M.~Rahaman,
``Possible existence of wormholes in the central regions of halos,''
Annals Phys. \textbf{350} (2014), 561-567
[arXiv:1501.00490 [physics.gen-ph]].

\bibitem{Rahaman:2013xoa}
F.~Rahaman, P.~K.~F.~Kuhfittig, S.~Ray and N.~Islam,
``Possible existence of wormholes in the galactic halo region,''
Eur. Phys. J. C \textbf{74} (2014), 2750
[arXiv:1307.1237 [gr-qc]].



\bibitem{Krishnendu:2017shb}
N.~V.~Krishnendu, K.~G.~Arun and C.~K.~Mishra,
``Testing the binary black hole nature of a compact binary coalescence,''
Phys. Rev. Lett. \textbf{119}, no.9, 091101 (2017)
[arXiv:1701.06318 [gr-qc]].

\bibitem{Cardoso:2016oxy}
V.~Cardoso, S.~Hopper, C.~F.~B.~Macedo, C.~Palenzuela and P.~Pani,
``Gravitational-wave signatures of exotic compact objects and of quantum corrections at the horizon scale,''
Phys. Rev. D \textbf{94}, no.8, 084031 (2016)
[arXiv:1608.08637 [gr-qc]].

\bibitem{Aneesh:2018hlp}
S.~Aneesh, S.~Bose and S.~Kar,
``Gravitational waves from quasinormal modes of a class of Lorentzian wormholes,''
Phys. Rev. D \textbf{97}, no.12, 124004 (2018)
[arXiv:1803.10204 [gr-qc]].

\bibitem{Roy:2019yrr}
P.~Dutta Roy, S.~Aneesh and S.~Kar,
``Revisiting a family of wormholes: geometry, matter, scalar quasinormal modes and echoes,''
Eur. Phys. J. C \textbf{80}, no.9, 850 (2020)
[arXiv:1910.08746 [gr-qc]].




\bibitem{Lobo:2007qi}
F.~S.~N.~Lobo,
``A General class of braneworld wormholes,''
Phys. Rev. D \textbf{75}, 064027 (2007)
[arXiv:gr-qc/0701133 [gr-qc]].

\bibitem{deLeon:2009pu}
J.~P.~de Leon,
``Static wormholes on the brane inspired by Kaluza-Klein gravity,''
JCAP \textbf{11}, 013 (2009)
[arXiv:0910.3388 [gr-qc]].
%
\bibitem{Wong:2011pt}
K.~C.~Wong, T.~Harko and K.~S.~Cheng,
``Inflating wormholes in the braneworld models,''
Class. Quant. Grav. \textbf{28}, 145023 (2011)
[arXiv:1105.2605 [gr-qc]].

\bibitem{Kar:2015lma}
S.~Kar, S.~Lahiri and S.~SenGupta,
``Can extra dimensional effects allow wormholes without exotic matter?,''
Phys. Lett. B \textbf{750}, 319-324 (2015)
[arXiv:1505.06831 [gr-qc]].

\bibitem{Banerjee:2019ssy}
A.~Banerjee, P.~H.~R.~S.~Moraes, R.~A.~C.~Correa and G.~Ribeiro,
``Wormholes in Randall-Sundrum braneworld,''
[arXiv:1904.10310 [gr-qc]].

\bibitem{Wang:2017cnd}
D.~Wang and X.~H.~Meng,
``Traversable braneworld wormholes supported by astrophysical observations,''
Front. Phys. (Beijing) \textbf{13}, no.1, 139801 (2018)
[arXiv:1706.06756 [gr-qc]].



\bibitem{Kaluza:1921tu}
T.~Kaluza,
``Zum Unit\"atsproblem der Physik,''
Sitzungsber. Preuss. Akad. Wiss. Berlin (Math. Phys. ) \textbf{1921}, 966-972 (1921)
[arXiv:1803.08616 [physics.hist-ph]].

\bibitem{Klein:1926tv}
O.~Klein,
``Quantum Theory and Five-Dimensional Theory of Relativity. (In German and English),''
Z. Phys. \textbf{37}, 895-906 (1926).

\bibitem{GSW}
M.S. Green, J.H. Schwarz, E. Witten, ``Superstring theory'', Cambridge University Press, Cambridge, U.K

\bibitem{Rubakov:1983bb}
V.~A.~Rubakov and M.~E.~Shaposhnikov,
``Do We Live Inside a Domain Wall?,''
Phys. Lett. B \textbf{125}, 136-138 (1983).

\bibitem{Gogberashvili:1998vx}
M.~Gogberashvili,
``Hierarchy problem in the shell universe model,''
Int. J. Mod. Phys. D \textbf{11}, 1635-1638 (2002)
[arXiv:hep-ph/9812296 [hep-ph]].

\bibitem{Gogberashvili:1998iu}
M.~Gogberashvili,
``Our world as an expanding shell,''
Europhys. Lett. \textbf{49}, 396-399 (2000)
[arXiv:hep-ph/9812365 [hep-ph]].

\bibitem{Furey:2015yxg}
C.~Furey,
Ph. D. thesis, University of Waterloo
``Standard model physics from an algebra?,''
[arXiv:1611.09182 [hep-th]].

\bibitem{Baez:2001dm}
J.~C.~Baez,
``The Octonions,''
Bull. Am. Math. Soc. \textbf{39} (2002), 145-205
[erratum: Bull. Am. Math. Soc. \textbf{42} (2005), 213]
[arXiv:math/0105155 [math.RA]].

\bibitem{Baez:n-cafe}
J.~C.~Baez,
``Can We Understand the Standard Model Using Octonions?''
The n-Category Cafe, March 31, 2021,
https://golem.ph.utexas.edu/category/2021/03/can\_we\_understand\_the\_standard\_1.html

\bibitem{Baez:2010ye}
J.~C.~Baez and J.~Huerta,
``Division Algebras and Supersymmetry II,''
Adv. Theor. Math. Phys. \textbf{15} (2011) no.5, 1373-1410
[arXiv:1003.3436 [hep-th]].

\bibitem{Furey:2018yyy}
N.~Furey,
``Three generations, two unbroken gauge symmetries, and one eight-dimensional algebra,''
Phys. Lett. B \textbf{785} (2018), 84-89
[arXiv:1910.08395 [hep-th]].

\bibitem{Furey:2018drh}
C.~Furey,
``$SU(3)_C\times SU(2)_L\times U(1)_Y\left( \times U(1)_X \right) $ as a symmetry of division algebraic ladder operators,''
Eur. Phys. J. C \textbf{78} (2018) no.5, 375
[arXiv:1806.00612 [hep-th]].

\bibitem{Gillard:2019ygk}
A.~B.~Gillard and N.~G.~Gresnigt,
``Three fermion generations with two unbroken gauge symmetries from the complex sedenions,''
Eur. Phys. J. C \textbf{79} (2019) no.5, 446
[arXiv:1904.03186 [hep-th]].


\bibitem{Randall:1999ee}
L.~Randall and R.~Sundrum,
``A Large mass hierarchy from a small extra dimension,''
Phys. Rev. Lett. \textbf{83}, 3370-3373 (1999)
[arXiv:hep-ph/9905221 [hep-ph]].

\bibitem{Randall:1999vf}
L.~Randall and R.~Sundrum,
``An Alternative to compactification,''
Phys. Rev. Lett. \textbf{83}, 4690-4693 (1999)
[arXiv:hep-th/9906064 [hep-th]].


\bibitem{Kar:1995jz}
S.~Kar, S.~Minwalla, D.~Mishra and D.~Sahdev,
``Resonances in the transmission of massless scalar waves in a class of wormholes,''
Phys. Rev. D \textbf{51}, 1632-1638 (1995).

\bibitem{Ellis:1973yv}
H.~G.~Ellis,
``Ether flow through a drainhole - a particle model in general relativity,''
J. Math. Phys. \textbf{14}, 104-118 (1973);
 Errata: J.Math.Phys. 15 (1974) 520.

\bibitem{Bronnikov:1973fh}
K.~A.~Bronnikov,
``Scalar-tensor theory and scalar charge,''
Acta Phys. Polon. B \textbf{4}, 251-266 (1973).



\bibitem{Dzhunushaliev:2009va}
V.~Dzhunushaliev, V.~Folomeev and M.~Minamitsuji,
``Thick brane solutions,''
Rept. Prog. Phys. \textbf{73}, 066901 (2010)
[arXiv:0904.1775 [gr-qc]].

\bibitem{Ghosh:2008vc}
S.~Ghosh and S.~Kar,
``Bulk spacetimes for cosmological braneworlds with a time-dependent extra dimension,''
Phys. Rev. D \textbf{80}, 064024 (2009)
[arXiv:0812.1666 [gr-qc]].


\bibitem{Sengupta:2021wvi}
R.~Sengupta, S.~Ghosh, M.~Kalam and S.~Ray,
``Wormhole on the Brane with Ordinary Matter: The Broader View,''
[arXiv:2105.11785 [gr-qc]].

\end{thebibliography}
\end{document}